\documentclass[sigconf, prologue, dvipsnames, table]{acmart}

\usepackage{algpseudocode}
\usepackage{algorithm}
\usepackage{graphicx}
\usepackage{textcomp}
\usepackage{booktabs}
\usepackage{listings}
\usepackage{bbding}
\usepackage{xcolor}
\usepackage{pgfplots}
\usepackage{multirow}
\usepackage{multicol}
\usepackage{amsmath}
\usepackage{centernot}
\usepackage[utf8]{inputenc}

\newcommand{\codename}{{\tt VirtAnalyzer}}
\newcommand{\codenametwo}{{ DeClassifier}}

\newcommand{\ayomide}[1]{{\authnote{ayomide}{#1}}}
\newcommand{\ignore}[1]{}

\renewcommand{\paragraph}[1]{\medskip\noindent{\bf{#1.}}}

\newcommand{\specialcell}[2][c]{%
	\begin{tabular}[#1]{@{}c@{}}#2\end{tabular}}

\begin{document}
\widowpenalty10000
\clubpenalty10000

\title{Devil is Virtual: Reversing Virtual Inheritance in C++ Binaries}

\author{Rukayat Ayomide Erinfolami}
\affiliation{%
	\institution{Binghamton University}
	}
\email{rerinfo1@binghamton.edu}

\author{Aravind Prakash}
\affiliation{%
	\institution{Binghamton University}
}
\email{aprakash@binghamton.edu}

\begin{abstract}
  Complexities that arise from implementation of object-oriented concepts in C++ such as virtual dispatch and dynamic type casting have attracted the attention of attackers and defenders alike. 
  Binary-level defenses are dependent on full and precise recovery of class inheritance tree of a given program. 
  While current solutions focus on recovering single and multiple inheritances from the binary, they are oblivious to virtual inheritance. Conventional wisdom among binary-level defenses is that virtual inheritance is uncommon and/or support for single and multiple inheritances provides implicit support for virtual inheritance. In this paper, we show neither to be true. 
  
  Specifically, (1) we present an efficient technique to detect virtual inheritance in C++ binaries and show through a study that virtual inheritance can be found in non-negligible number (more than 10\% on Linux and 12.5\% on Windows) of real-world C++ programs including Mysql and libstdc++. (2) we show that failure to handle virtual inheritance introduces both false positives and false negatives in the hierarchy tree. These false positves and negatives either introduce attack surface when the hierarchy recovered is used to enforce CFI policies, or make the hierarchy difficult to understand when it is needed for program understanding (e.g., during decompilation). (3) We present a solution to recover virtual inheritance from COTS binaries. We recover a maximum of 95\% and 95.5\% (GCC -O0) and a minimum of 77.5\% and 73.8\% (Clang -O2) of virtual and intermediate bases respectively in the virtual inheritance tree.
\end{abstract}


\settopmatter{printacmref=false}
\renewcommand\footnotetextcopyrightpermission[1]{}
\maketitle

\pagestyle{empty}
\section{Introduction}\label{sec:intro}
Recovering high-level semantic information from binaries has strong security relevance in areas such as vulnerability detection, control-flow integrity (CFI)~\cite{zhang2016vtrust, norax, ravel, Dewey:2012:CFI, secret, gawlik:2014:tvip, vantough:call:2016, vtpin}, decompilation~\cite{bao2014byteweight, eissec, eklavya, multiverse, uroboros, ramblr} and memory forensics~\cite{origen}. 
In particular, recovery of object-oriented semantics (e.g., class hierarchy) is key to C++ binary-level defenses (e.g., ~\cite{declassifier,fan2017vip,gawlik:2014:tvip,prakash:2015:vfguard,marx:andre}). 
\ignore{Reverse engineering C++ high-level semantics present significant challenges. First, due to the underlying complexities arising from the implementation of inheritance and polymorphism, seemingly simple and straightforward source-level constructs translate into complex code and inter-dependent data.
	Second, due to the high prevalence of optimization, some commonly-relied-upon structural elements (e.g., RTTI) may be absent. 
	Finally, implementation-level flexibilities exercised by different compilers in implementing polymorphism make heuristic approaches infeasible.}

Traditional C++ binary analysis solutions have focused on constructor analysis~\cite{vci:2017,Dewey:2012:CFI}, destructor analysis~\cite{fokin:2011:C++Rev}, overwrite analysis~\cite{marx:andre}, and VTable analysis~\cite{prakash:2015:vfguard,zhang:2015:vtint,gawlik:2014:tvip} in order to recover at least a partial class hierarchy tree (CHT). 
While prior solutions have focused on recovering single and multiple inheritances in the binary, virtual inheritance---an important feature of C++ language---has been ignored. 
From a security standpoint, some key questions arise: 1) How common is virtual inheritance? 2)  Is virtual inheritance relevant for security? 3) Does support for single and multiple inheritance implicitly cover virtual inheritance? 

\vspace{.04in}
\noindent 
{\bf Virtual inheritance is not uncommon:}
Virtual inheritance in C++ facilitates implementation of key design ideas, and has been used in prominent and widely-used programs (e.g., libstdc++, Mysql). 
Our first study comprising of 1129 Linux C++ binaries found 11\% of the libraries to contain virtual inheritance while our second study of 648 Windows binaries found 12.5\% with virtual inheritance.  \ignore{In fact, out of 219 C++ libraries, 26, i.e., over 10\% of libraries contain virtual inheritance.} 
Widely-used libraries such as libstdc++ utilize virtual inheritance to prevent duplication of stream objects in the IO-related classes. Because libstdc++ is linked to all C++ programs, virtual inheritance can be commonly found in most C++ programs' memory. 

\vspace{.04in}
\noindent
{\bf Security relevance of virtual inheritance:}
Failure to handle virtual inheritance results in severe security flaws. 
Current binary-level CFI defenses against C++ virtual dispatch attacks extract the VTables in a binary and given a callsite, they construct a policy that allows the callsite to target a strict subset of polymorphic virtual functions derived from the class inheritance tree. 
Without specific mechanisms to handle virtual inheritance, current solutions either suffer from false negatives or false positives in the inheritance tree. In case of Marx~\cite{marx:andre}, compiler-generated ``construction VTables" (transient VTables used in construction of objects with virtual bases) are incorrectly included in the inheritance tree as regular VTables (i.e., VTables that represent a class) thereby resulting in false positives in the inheritance tree.  Whereas in the case of VCI~\cite{vci:2017} legitimate inheritance relationships arising due to virtual inheritance are completely missed due to the lack of support for virtual inheritance. VCI is testament to the fact that {\em support for single and multiple inheritance does not implicitly cover virtual inheritance.}
Both false positives and negatives result in inaccuracies in resulting CFI policies.

Unlike single and multiple inheritances, recovery of virtual inheritance poses significant technical challenges. First, thanks to the {\em is-a} property, reference to a derived object is also a legitimate reference to its virtual base object. However, by definition, a single copy of the virtual base is retained in the entire inheritance tree. 
As such, offset of the virtual base subobject from a derived class object and an intermediate class subobject (i.e., object of a class between the derived class and the virtual base class in the inheritance tree) could be different. 
Any binary-level static object-layout analysis that intends to capture virtual bases must take into account various offsets from different derived objects in an inheritance tree. 
Second, the ABI~\cite{ItaniumABI} necessitates additional structures and fields, e.g., virtual base offset (vbase-offset), virtual call offset (vcall-offset), construction VTables, Vbtable, etc. in order to implement virtual inheritance. 
These fields and structures introduce complexities in implementation that require special handling. Finally, virtual bases are allocated at the end of all the non-virtual bases in an inheritance (sub) tree. It is therefore important for a virtual inheritance recovery solution to delineate between non-virtual and (one or more) virtual bases in an object's memory. 

In this paper, we first show that virtual inheritance is not uncommon. To this end, we perform a comprehensive study of C++ binaries in the default installation of Ubuntu Linux 18.04 distribution and report that 11\% of C++ libraries contain virtual inheritance. We also performed a study of Windows C++ DLLs and report that 12.5\% of them contain virtual inheritance.
Further, we design a robust virtual inheritance recovery engine that pivots on the ABI definitions (both Itanium for gcc and clang, and MSVC for Microsoft Visual Studio). 
Our solution is tolerant to compiler variations including optimizations. 
Our class inheritance engine codenamed \codename\ employs object-offset analysis that can identify virtual bases in a derived object with a high level of precision. 
\codename\ is able to successfully recover a maximum of 95\% and 95.5\% (GCC -O0) and a minimum of 77.5\% and 73.8\% (Clang -O2) of virtual and intermediate bases respectively in the virtual inheritance tree. 


Our contributions can be summarized as follows: 
\begin{enumerate}
	\item We present simple and efficient algorithms to detect presense of and recover virtual inheritance in a given C++ binary that adheres to either Itanium or MSVC ABI. Our techniques are ABI-based, and so are largely unaffected by the specific compiler and/or optimizations (except in cases where entire classes are removed by the compiler). 
	\item We show that virtual inheritance is not uncommon in C++ binaries with significant security concerns. It cannot be ignored.
	\item We presented a sample attack that depicts how false positives in the CHT due to virtual inheritance can be exploited despite state-of-the-art defenses. We further demonstrate that an exponential (O($n^2$)) attack surface manifests where $n$ is the depth of the inheritance subtree with a virtual base. 
	
\end{enumerate}

\ignore{
	\vspace{.08in}
	\noindent
	{\bf Paper Organization:} In Section~\ref{sec:background} we present the technical background to our work. We describe the security impact of virtual inheritance in Section~\ref{sec:security_impact}. Section~\ref{sec:overview} presents the overview of our work while Section~\ref{sec:design} presents the details of \codename. We present implementation details and evaluation in Section~\ref{eval} and related work in Section~\ref{sec:related}. Finally, we conclude in Section~\ref{sec:conclude}.
}

\section{Technical Background}\label{sec:background}
In this section, we provide the technical details needed to understand the remainder of the paper. Because the Itanium ABI is widely used (adhered to by gcc and clang) and ABI is openly available~\cite{ItaniumABI},  we use it as a focal point of our work. However, our work also supports MSVC ABI.

\subsection{Running Example}
We will use the running example in Listing~\ref{virtualinh} throughout the paper. Class A is inherited virtually by each of classes B and C. Class D inherits from classes B and C to form what is popularly known as the ``diamond" structure. 
Because class A is inherited virtually, only one copy of the subobject of A is retained in the object of class D. Listing~\ref{l3} shows the disassembly of D's constructor. All code examples in this paper were compiled using GCC 7.3 with optimization flag O0, except otherwise stated

\lstset{language=[x86masm]Assembler, commentstyle=\color{Gray}, keywordstyle=\color{MidnightBlue}, basicstyle=\footnotesize\ttfamily, identifierstyle=\color{RedViolet}}

\lstset{language=Java}
\begin{lstlisting}[belowskip=-\baselineskip,caption={Running example}, label={virtualinh}, float,floatplacement=H, multicols=2]
class A{
public:
int a;
virtual void af(){...}
};
class B: public virtual A{
public:
int b;
virtual void bf(){...}
};
class C: public virtual A{
public:
int c;
virtual void cf(){...}
};
class D: public B, public C{
public:
int d;
virtual void df(){...}
};
\end{lstlisting}

\lstset{language=[x86masm]Assembler}
\begin{lstlisting}[aboveskip=-\baselineskip, belowskip=-\baselineskip, caption={Disassembly of the constructor of D in the running example.}, label={l3}, float,floatplacement=H]
...
1.  mov [rbp+var_8], rdi
2.  mov rax, [rbp+var_8]
3.  add rax, 20h
4.  mov rdi, rax; this, at offset 20h
5.  call _ZN1AC2Ev; A::A(void)
...
6.  mov rax, [rbp+var_8]
7.  lea rdx, off_201BB8; subVTT address of B-in-D
8.  mov rsi, rdx
9.  mov rdi, rax; this, at offset 0
10. call _ZN1BC2Ev; B::B(void) primary base class
...
11. mov rax, [rbp+var_8]
12. add rax, 10h
13. lea rdx, off_201BC8; subVTT address of C-in-D
14. mov rsi, rdx
15. mov rdi, rax; this, at offset 10h
16. call _ZN1CC2Ev; C::C(void)
...
\end{lstlisting}

\ignore{\em Definitions of key terms used in this paper are reproduced in Appendix~\ref{appendix:terms} for reviewers' convenience.}

	\subsection{Key Definitions}
	\vspace{.06in}
	\noindent
	{\bf Polymorphic class:} is a class that declares at least a virtual function or derives directly or indirectly from a class that is polymorphic. 
	
	\vspace{.03in}
	\noindent
	{\bf Direct and Indirect base:} Class DB is said to be the direct base of class C if C inherits directly from DB. Whereas, class IB is said to be an indirect base of class C if there exists at least one class M such that M inherits from IB and C inherits from M. 
	
	\vspace{.03in}
	\noindent
	{\bf Primary and Secondary VTables:} Primary VTable of a class C contains the virtual functions defined in C. It is shared with C's primary base. C has a secondary VTable associated with each of its secondary bases. The secondary VTables of a class are laid out immediately after the primary VTable, as a result, the address of a secondary VTable is always greater than that of its primary VTable.
	
	\vspace{.03in}
	\noindent
	{\bf Object and Subobject:} An object of class C contains entries for vptrs to C's VTables and entries for all non-static member variables of C. A subobject in C's object belong specifically to C or one of its base classes and it contains a vptr and non-static member variables defined in C or the base class. For instance, Figure~\ref{fig:obj_layout} shows that C's object contains two subobjects.
	
	\vspace{.03in}
	\noindent
	{\bf Virtual Inheritance Tree:} This a subtree in the class hierarchy tree rooted at a virtual base.
	
	\vspace{.03in}
	\noindent
	{\bf Virtual Call Offset}
	Every virtual function defined in the virtual base class has a vcall-offset entry in the secondary VTable (of the derived) corresponding to the virtual base. Since the virtual base could be shared among multiple base classes of a derived class (e.g. B and C in the running example), there is the need to identify the derived class with the most recent definition. The associated vcall-offset is equal to the offset of the virtual base subobject from the derived subobject with the most recent definition. Functions which are not overriden by a derived class have vcall-offset of zero, while the others have negative vcall-offsets.

\ignore{
	\subsection{Multiple Inheritance}
	C++ allows a class to inherit from more than one classes. This case of inheritance is referred to as multiple inheritance. The derived class inherits all the member variables and member functions defined in all its base classes. With polymorphism a derived class can override functions defined by its base classes. The compiler generates a complete object VTable for each polymorphic class in a given program.
}

\subsection{Virtual Inheritance}
Virtual inheritance is the solution to the ``diamond" structure problem, wherein multiple inheritance results in multiple copies of a base class' member variable(s) in the object of a derived class. In Listing~\ref{virtualinh}, since B and C virtually inherit from A, the compiler is instructed to keep a single copy of A in D. The object of D is such that there is exactly one copy of A's subobject, which is placed at the end of D's object. Virtual inheritance is achieved by prefixing the base class name in the class signature with the keyword ``virtual".

\begin{figure}
	\centering
	\includegraphics[scale=0.4]{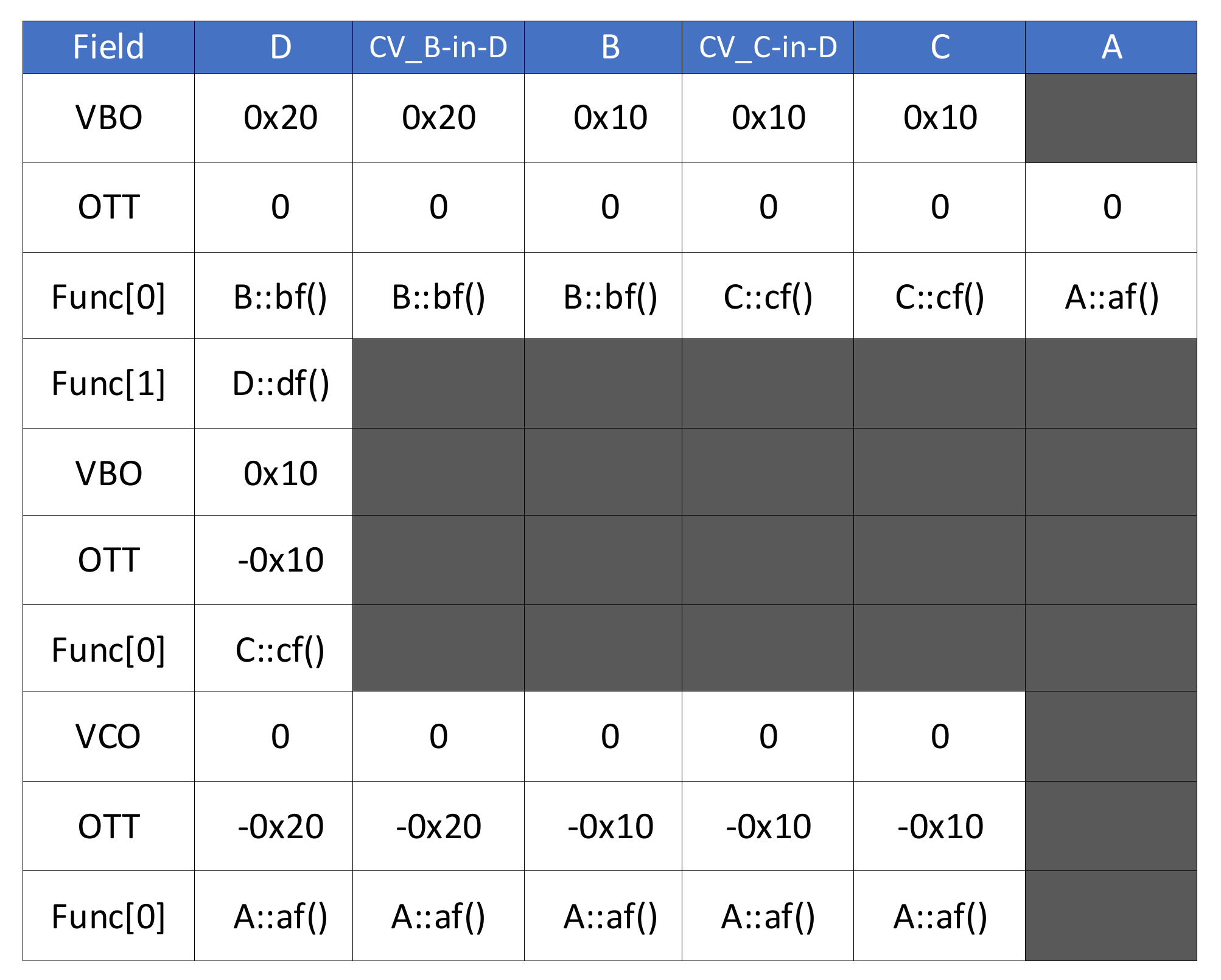}
	\caption{VTable fields of classes in the running example. ``CV" means construction VTable, VBO means vbase-offset, OTT means offset-to-top}\label{fig:vtable_layout}
\end{figure}

\vspace{.06in}
\noindent{\bf Virtual Base Offset}\label{vbaseoffset}
Every class which inherits from a virtual base, either directly or indirectly, has a ``vbase-offset". It is the offset of a virtual base subobject from a derived object. This value is used when a member variable in the virtual base subobject needs to be accessed from a pointer pointing to a derived object. It is also used during the initialization of the secondary VTable corresponding to a virtual base. The VTable of a class has a vbase-offset field for each of its virtual bases. For instance, as shown in Figure~\ref{fig:vtable_layout}, D has two vbase-offsets of values 0x20 (offset from D's subobject to A-in-D's subobject) and 0x10 (offset from C-in-D's subobject to A-in-D's subobject).

\ignore{
	\subsubsection{Virtual Call Offset}\label{vcalloffset}
	Every virtual function defined in the virtual base class has a vcall-offset entry in the secondary VTable (of the derived) corresponding to the virtual base. Since the virtual base could be shared among multiple base classes of a derived class (e.g. B and C), there is the need to identify the derived class with the most recent definition. The associated vcall-offset is equal to the offset of the virtual base subobject from the derived subobject with the most recent definition. Functions which are not overriden by a derived class have vcall-offset of zero, while the others have negative vcall-offsets.
}

\ignore{
	\subsubsection{Virtual Thunk}
	The virtual thunk adjusts the \texttt{this} pointer to point to the subobject with the most recent definition of the function it is associated with. It does this by obtaining the vcall-offset of that function from the secondary VTable corresponding to the virtual base and then adding it to the \texttt{this} pointer passed to it.
}

\vspace{.06in}
\noindent{\bf Construction VTable}\label{cons_VTable}
Construction VTables are used during construction and destruction of intermediate bases in a virtual inheritance tree. They are needed to access the correct vbase-offset and virtual functions associated to a given base. Consider the running example and Fig~\ref{fig:vtable_layout}, B needs to be constructed in D. If B's VTable is used for the construction of B in D, the vbase-offset that will be retrieved is 0x10 (offset to A's subobject in B). However, the vbase-offset of D (shared with B) is 0x20.  Retrieving a vbase-offset of 0x10 instead of 0x20 will result in accessing the wrong location in D's object as A's subobject. Therefore, there is the need for another VTable corresponding to B-in-D that has the correct vbase-offset(0x20) and the virtual functions associated with B (this is because while constructing B's subobject in D, B's virtual functions should be accessible not those of D), as well as special constructors and destructors which access the construction VTables.

Figure~\ref{fig:vtable_layout} shows the fields of the two construction VTables in the running example (CV\_B-in-D, CV\_C-in-D). Every intermediate base class has an associated construction VTable. As shown in Table~\ref{VTable-and-CVTable}, the construction VTable of an IntermediateBase-in-Derived has the vbase-offset, vcall-offset and offset-to-top of virtual bases associated with the derived, while the virtual functions, type info and offset-to-top of non-virtual bases associated with the intermediate base.

\begin{table}[]
	\centering
	\caption{Differences and similarities among the fields of a VTables and a Construction VTable using running example}
	\label{VTable-and-CVTable}
	\rowcolors{1}{white}{gray!30}
	\scalebox{0.8}{%
		\begin{tabular}{lcc}
			\toprule[0.4ex]
			\textbf{Fields} & \textbf{\specialcell{Construction \\VTable of B-in-D}} & \textbf{VTable of B} \\ \midrule
			vbase-offset & \specialcell{higher e.g 0x20\\ in B-inD} & \specialcell{lower e.g \\0x10 in B} \\ 
			\specialcell{offset-to-top \\(non-virtual \\subVTable)} & same & same \\ 
			\specialcell{offset-to-top \\(for virtual \\subVTable)} & \specialcell{lower e.g -0x20\\ for A's VTable\\ in B-in-D} & \specialcell{higher e.g -0x10\\ for A's VTables\\ in B} \\ 
			vcall-offset & \specialcell{lower e.g -0x20\\ for af() in B-in-D} & \specialcell{higher e.g \\-0x10} \\ 
			type-info & Same & Same \\ 
			Function pointers & Same & Same \\ \bottomrule[0.4ex]
		\end{tabular}
	}
\end{table}

\begin{figure}[ht]
	\centering
	\includegraphics[scale=0.4]{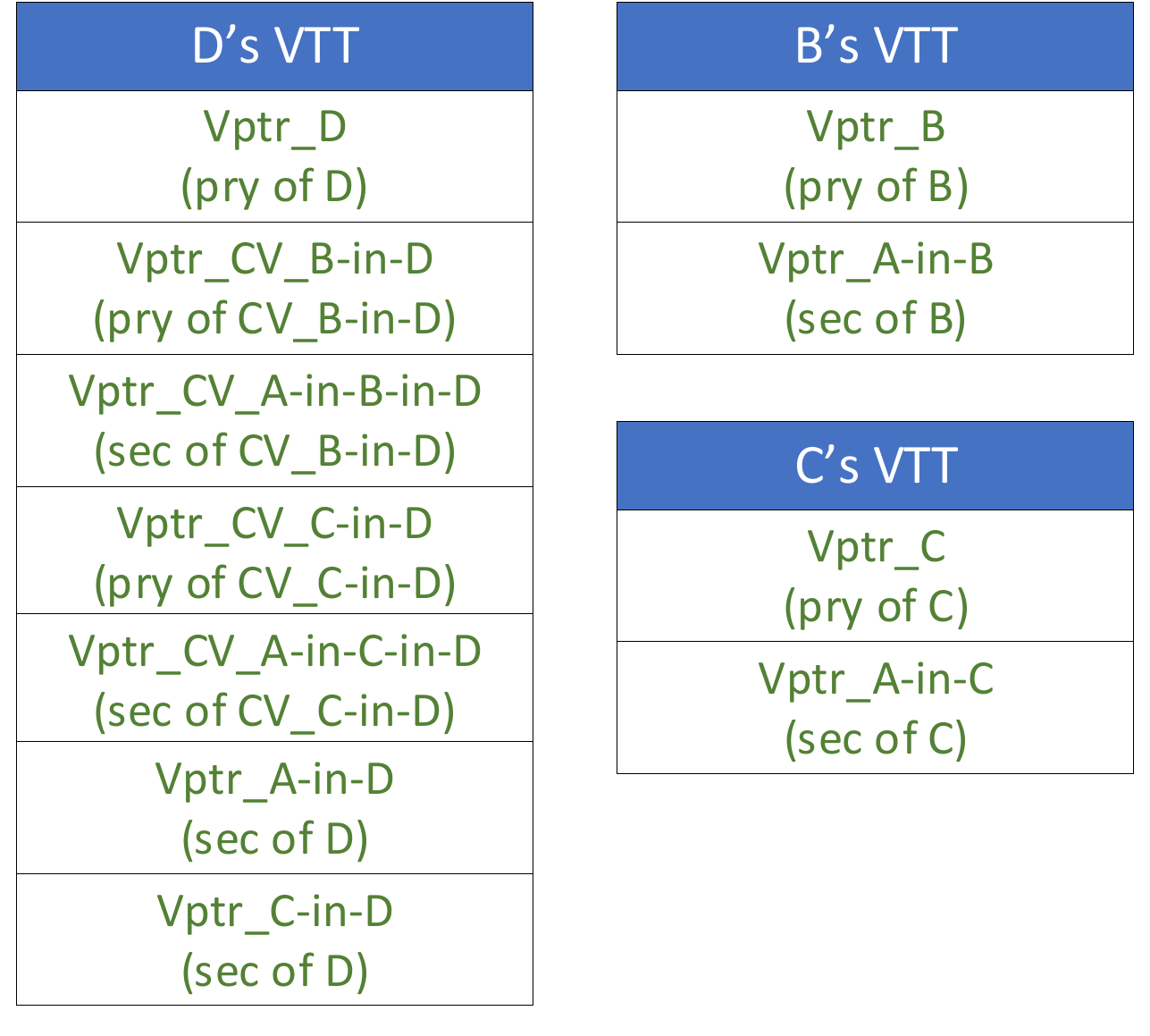}
	\caption{The VTT layout of classes in running example.``CV" means construction VTable, ``pry" means primary VTable, ``sec" means secondary VTable}\label{fig:vtt}
	
\end{figure}

\vspace{.06in}
\noindent{\bf Virtual Table Table}\label{vtt}
The virtual table table (VTT)\footnote{VCI~\cite{vci:2017} uses the acronym VTT to refer to VTable group, which is different from the Virtual Table Tables defined in the Itanium ABI~\cite{ItaniumABI}. In this paper, we stick to the terminology used in the ABI.} is an array of VTable pointers and construction VTable pointers (if any exist) of a class (Figure~\ref{fig:vtt}). Every derived class with at least one direct or indirect virtual base class(es) has an associated VTT. The VTT is made up of pointers to the primary and seconday VTables of the derived class, and the primary and secondary construction VTables of its intermediate base classes. We refer to each complete object VTable pointers (i.e. group of primary and secondary VTable pointers of a class) within a VTT as SubVTTs. Basically, a VTT is made up of multiple subVTTs each pointing to a complete object VTable or a construction VTable. Only one of those subVTTs point to the VTable of the derived class which owns the VTT, while the others point to construction VTables. 

As mentioned in Section~\ref{cons_VTable}, there is need for special constructors (and destructors) to construct intermediate base subobjects in the derived object. The constructor of the derived class passes a pointer to the subVTT corresponding to the construction VTable of the IntermediateBase-in-Derived to the special constructor as a second hidden argument. It then initializes the intermediate base subobject by accessing the subVTT (Listing~\ref{l3}, lines 7 and 13) as opposed to initializing with immediate values in the case of single and multiple inheritance. \ignore{Apart from the base class subobject address, the derived class constructor also passes the subVTT address as the second argument to this special constructor of the base class.}

\vspace{.04in}
\noindent{\bf Order of Object Construction}
We will explain this using the running example and Listing~\ref{l3}. First, the constructor of D constructs all its virtual bases, A in this case. Next B is constructed by calling the special constructor of B with the address of the subVTT corresponding to the construction VTable B-in-D as the second argument. The same is done for C. Finally, the vptrs of D are written into appropriate locations in the object starting with the primary vptr.

\begin{figure}
	\centering
	\includegraphics[height=5cm, width=9cm]{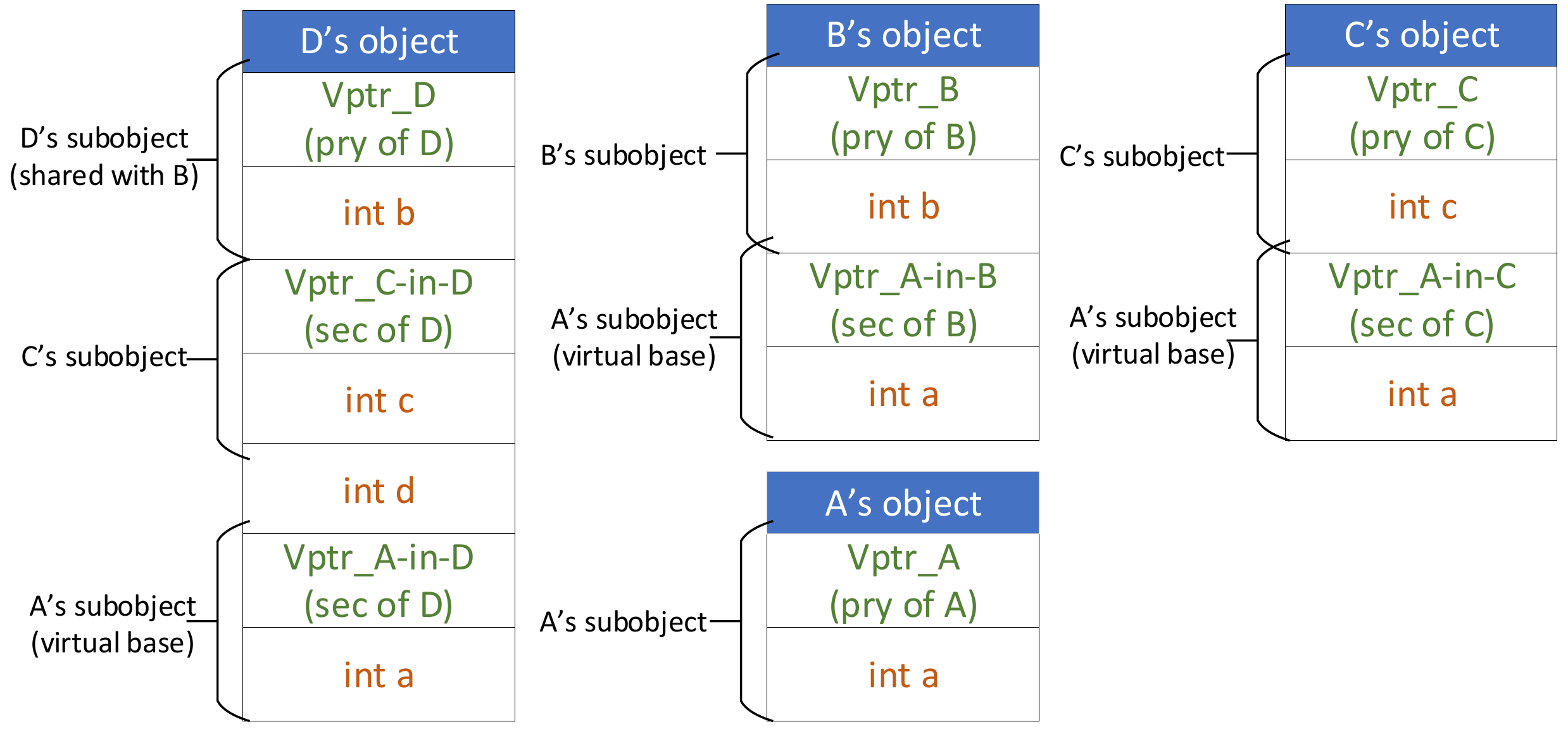}
	\caption{Object layout of classes in the running example. mark the subobjects, include VTTs}\label{fig:obj_layout}
\end{figure}

\vspace{.04in}
\noindent{\bf Object and VTable Layout with Virtual Inheritance}
Figure~\ref{fig:obj_layout} shows objects of types A, B, C and D. An object of D also contains a subobject of A. Note that none of the objects share subobject with A, since the subobject of A could be shared among multiple bases. This is also the same for VTables Figure~\ref{fig:vtable_layout}, a derived class does not share VTable with its virtual base (except the virtual base is either empty or nearly empty). Note that only the VTable corresponding to the virtual base has vcalloffset fields.

\section{Security Impact of Virtual Inheritance}\label{sec:security_impact}
\subsection{Study: Virtual Inheritance in Real-World Programs}
Virtual inheritance in C++ has received very little attention.
Prior efforts have focused on single and multiple inheritances, therefore, support for handling virtual inheritance is missing in both source-code-level solutions\cite{jang:2014:CFI:Src, vtvgcc:2012:tice, haller2015shrinkwrap, fan2017vip, zhang2016vtrust} and binary-level solutions\cite{marx:andre, vci:2017, fokin:2011:C++Rev, katz:2016, lego14}. 
While it is true that virtual inheritance in C++ is not as common as single or multiple inheritance, we conducted a study in order to understand how (un)common virtual inheritance is in realworld programs. 
We evaluated 1129 C++ binaries on Ubuntu 18.04 Linux distribution and found 11\% of the libraries with instances of virtual inheritance ranging from 1 to 27. For Windows distribution, we evaluated 648 DLLs and found 12.5\% with instances of virtual inheritance ranging from 1 to 382. Our approach for detecting virtual inheritance in a binary is described in detail in Sections~\ref{subsec:extraction} and ~\ref{subsec:recovery}. 
Findings of our study are tabulated in Table~\ref{vh_distr}. Notably, we found that virtual inheritance is prevalent in both libraries and executables including the libstdc++ library and mysql database engine. 

\ignore{
	We started with a total of 2097 libraries and 22901 executables that included both C++ and non-C++ binaries. We obtained the libraries by gathering all preinstalled libraries on Ubuntu. The executables were obtained by running a script that uses Linux's ``apt" command to identify programs with at least one .cpp file. To filter out non-C++ programs, we searched each binary for ABI specific name mangling information (i.e., {\tt cxxabi}). After filtering for C++ binaries, we ended up with a total of 219 libraries and 909 executables. Table \ref{vh_distr} shows these numbers as well as the number of binaries with virtual inheritance (11\% libraries including libstdc++ and 2\% executables including Mysql). 
}
State-of-the-art binary analysis tools that rely on inference of class hierarchy like Marx\cite{marx:andre} and VCI\cite{vci:2017} do not recover virtual inheritance, which is necessary for enforcing precise CFI policies.

\begin{table}[ht]
	\centering
	\normalsize
	\caption{Prevalence of virtual inheritance in C++ programs}
	\label{vh_distr}
	\scalebox{0.8}{%
		\begin{tabular}{llcc}
			\toprule[0.4ex]
			ABI&& \# C++ & \# with Virtual Inheritance \\ \midrule
			\multirow{2}{*}{Itanium} &Libraries &  219 & 26 \\
			&Executables &  910 & 19  \\
			MSVC&DLLs &  648 & 81 \\ \bottomrule[0.4ex]
			
		\end{tabular}
	}
\end{table}

\subsection{False Positives and False Negatives in State-of-the-art Binary Level Solutions}\label{sec-impl}
We consider state-of-the-art binary analysis tools which reconstruct high level semantics from the binary. Table~\ref{bin_analy_solns} shows the weaknesses (introduction of either false positives or false negatives) of the solutions in handling virtual inheritance. As representative solutions, we also provide a detailed description of  how VCI~\cite{vci:2017}, SmartDec~\cite{fokin:2011:C++Rev} and MARX~\cite{marx:andre} behave when virtual inheritance is present in a binary.

\begin{table}[ht]
	\centering
	\small
	\caption{Binary level solutions which recover high level semantic information from the binary}
	\label{bin_analy_solns}
	\scalebox{0.8}{%
		\begin{tabular}{lccc}
			\toprule[0.4ex]
			Solution & \specialcell{Introduces\\False +ve} & \specialcell{Introduces\\False -ve} & \specialcell{Distinguishes\\ Construction\\ VTables from\\ Regular VTables} \\ \midrule
			VCI\cite{vci:2017} & \XSolidBrush & \Checkmark & \XSolidBrush \\
			Marx\cite{marx:andre} & \Checkmark & \XSolidBrush & \XSolidBrush \\
			SmartDec\cite{fokin:2011:C++Rev} & \XSolidBrush & \Checkmark & \XSolidBrush \\
			OOAnalyzer\cite{ooanalyzer:2018} & \XSolidBrush & \XSolidBrush & \XSolidBrush \\
			vfGuard\cite{prakash:2015:vfguard} & \Checkmark & \XSolidBrush & \XSolidBrush \\
			Katz et al.\cite{katz:2016}& \Checkmark & \XSolidBrush &  \XSolidBrush \\
			ROCK\cite{Katz:2018} & \Checkmark & \XSolidBrush &  \XSolidBrush \\
			ObjDigger\cite{Jin:2014:objdigger}& \Checkmark & \XSolidBrush &  \XSolidBrush \\
			Lego\cite{lego14}& \Checkmark & \XSolidBrush &  \XSolidBrush \\
			Hex Rays\cite{igor:ida:decompiler}& \XSolidBrush & \Checkmark &  \XSolidBrush \\ 
			BinCFI\cite{Zhang:2013:CFI:Bin}& \Checkmark & \XSolidBrush &  \XSolidBrush \\
			VTint\cite{zhang:2015:vtint}& \Checkmark & \XSolidBrush &  \XSolidBrush \\
			Our Analysis & \XSolidBrush & \XSolidBrush & \Checkmark \\ \bottomrule[0.4ex]
			
		\end{tabular}
	}
\end{table}

\paragraph{False Positives in Marx}\label{marx-error}

{\em Marx Overview:} Marx is a binary-level solution that defends against abuse of virtual dispatch mechanism in C++. In a nutshell, for a given C++ virtual callsite, Marx identifies all the polymorphic functions that can be invoked at that callsite, and instruments the binary to allow only those functions. Allowable polymorphic functions are recovered by performing overwrite analysis which identifies sets of vptrs that get overwritten in an object during construction or destruction. Marx groups classes into sets wherein each set represents a class inheritance sub-tree with no particular inheritance order. Detailed description of Marx can be found in the paper~\cite{marx:andre}. 

For the running example, Marx recovers six complete-object VTables (including construction VTables). One VTable each for A, B, C and D and one construction VTable each for B-in-D and C-in-D. 
Marx does not make any distinction between VTables and construction VTables. 
In other words, Marx will interpret construction VTables to be representations of legitimate classes in the binary. 
For the running example, 12 VTables (breaking them into primaries and secondaries) are recovered. Under O0 optimization, Marx groups all 12 into a single set, while for O2, there are 3 sets. The sets for O2 are: 

\ignore{
	\begin{itemize}
		\item set1 = \{$B-in-D_{pry}(c), B_{pry}(a), D\_{pry(a shared with B)}$\}
		\item set2 = \{$C-in-D_{pry}(c), C_{pry}(a), D_{sec}(a corr. to C)$\}
		\item set3 = \{$B-in-D_{sec}(c corr. to A) ,B_{sec}(a corr. to A), C-in-D_{sec}(c corr. to A), C_{sec}(a corr. to A),  D_{sec}(a corr. to A), A_{pry}$\}
	\end{itemize}
}
\begin{itemize}
	\item set1 = \{$CV\_B-in-D_{pry}, B_{pry}, D_{pry}$\}
	\item set2 = \{$CV\_C-in-D_{pry}, C_{pry}, D_{sec}$\}
	\item set3 = \{$CV\_B-in-D_{sec} ,B_{sec}, \\CV\_C-in-D_{sec}, C_{sec}, D_{sec}, A_{pry}$\}
\end{itemize}

Set1 shows that Marx will incorrectly allow the vptr of Construction VTable B-in-D at a callsite that expects an object of type B. Consider listing~\ref{main_func} and listing~\ref{bf} which show a call to function B::bf() and the disassembly of function B::bf() respectively. B::bf() simply adds up member variable b of class B with the member variable a of class A and writes the result back to a. If an attacker successfully overwrites the vptr of the object of type B with that of the construction VTable B-in-D before the callsite, Marx would not raise any alarm. 

We see from Table~\ref{VTable-and-CVTable} that the vbase-offset in the construction VTable B-in-D is greater than that in the VTable of B. If the construction VTable of B-in-D is used, line 4 of listing~\ref{bf} retrieves a vbase-offset equal to 0x20 which is used to access the virtual base subobject at line 7. This is clearly outside the bounds of B's object since the total size of B is 0x20. Then lastly, an offset of 0x28 (0x20 + 0x8) from B's object is accessed at line 8 to get a member of the virtual base, and written in line 10. These read and write operations occur outside B's object bounds. The failure of Marx to handle virtual inheritance (meaning that construction VTables are neither identified nor filtered out from inheritance sets) introduces an attack surface for data corruption and arbitrary code execution under favorable circumstances. A detailed PoC attack is provided in Section~\ref{poc_attack}.

\lstset{language=C++}
\begin{lstlisting}[belowskip=-2\baselineskip, caption={Source code for main function}, label={main_func}, float,floatplacement=H]
1. int main(){
2. 		B *b = new B();
3. 		b->bf();
4. 		return 0;
5. }
\end{lstlisting}

\lstset{language=[x86masm]Assembler}
\begin{lstlisting}[belowskip=-2\baselineskip, caption={Disassembly for function B::bf()}, label={bf}, float,floatplacement=H]
...
1. mov rax, [rbp+var_8] ; get object address
2. mov ecx, [rax+8]; get member at offset 0x8
3. mov rax, [rax] ; get object's VTable
4. sub rax, 18h ; locate vbase-offset field
5. mov rdx, [rax] ; get vbase-offset
...
6. mov rax, [rbp+var_8] ; get object addres
7. add rax, rdx ;locate virtual base (VB) subobject
8. mov edx, [rax+8]; get virtual base member
...
9. add edx, ecx 
10. mov [rax+8], edx ; write result to VB subobject

\end{lstlisting}

\paragraph{False Negatives in VCI and SmartDec}

{\em VCI and SmartDec Overview:} VCI and SmartDec are binary level class hierarchy recovery tools which attempt to reason about the direction of inheritance. VCI achieves this by performing constructor only analysis, while SmartDec performs constructor and destructor analysis. Both solutions analyze the order in which base class constructors and/or destructors are called from the derived class constructor/destructor. The analysis is done by simply scanning constructors, for instance, for assembly callsites that invoke other constructors. While VCI puts measures in place to filter out composed classes from the class hierarchy tree, SmartDec includes both composed and inherited classes in the class hierarchy tree.

Regular constructors are known to be functions which initialize vptrs as immediate values. As mentioned in subsection~\ref{vtt}, special constructors used to construct intermediate bases in a virtual inheritance tree do not initialize any vptrs using immediate values. \ignore{To describe this problem further, let us add a main function (source code in Listing~\ref{main_func}, disassembly in Listing~\ref{main_asm}) to our running example.} Considering the disassembly in Listing~\ref{l3}, VCI and SmartDec will recover A as the direct base of D and ignore B and C since they are unable to identify any calls to the constructors of B and C from the constructor of D. VCI keeps a metadata in the binary which maps every function to the set of class types it can be invoked on, therefore for each function in the example, a map that looks like this is generated: af() = \{A,B,C,D\}, bf() = \{B\}, cf() = \{C\}, df() = \{D\}. Say function bf() is to be invoked on an object of type D as shown in Listing~\ref{main_asm}, before the callsite at line 10, VCI checks if there is a class in the set for bf() which has a vptr equal to the vptr obtained at line 6. Since the vptr belong to D and D is not in the set, VCI raises a false violation alarm. SmartDec will behave similarly. 

\lstset{language=[x86masm]Assembler}
\begin{lstlisting}[aboveskip=-0.5\baselineskip, belowskip=-1.8\baselineskip, caption={Disassembly for main function}, label={main_asm}, float,floatplacement=H]
1.  call _Znwm
2.  mov rbx, rax
...
3.  call _ZN1DC1Ev; D::D(void)
4.  mov [rbp + var_18], rbx
5.  mov rax, [rbp+ var_18]
6.  mov rax, [rax]; deref this ptr to get vptr
7.  mov rax, [rax]; deref vptr to get addr of bf()
8.  mov rbx, [rbp+ var_18]
9.  mov rdi, rax
10. call rax

\end{lstlisting}

\section{Exploiting Virtual Inheritance}\label{sec:exploit}
\subsection{Defeating Marx}\label{poc_attack}
In this section we present a proof-of-concept attack launched against a synthetic vulnerable program, Listing~\ref{attack}. The victim program is hardened with an implementation of Marx VTable protection policy (only Marx hierarchy recovery tool is open source, not the VTable protection tool). This policy ensures that only virtual functions from the set of classes related to the callsite type are allowed at runtime. The attack is successful because Marx does not differentiate between regular and construction VTables. \ignore{In the next subsection, we will describe the attack model as well as the attack.}

\vspace{.04in}
\noindent{\bf Attack Model}
The objective of this attack is to execute arbitrary code. We assume that the attacker can bypass ASLR and stack protector.

The first assumption allows the attacker to identify the absolute addresses of  suitable construction VTables to use in the attack. Bypassing ASLR to reveal such information is possible as shown in the literature~\cite{dmitry2016,ralf2013}. Once the address of a suitable construction VTable has been found, there is a need to write it into appropriate location in the object. Our PoC exploits buffer overflow vulnerability for this which is possible by bypassing stack protector. There are works that show that bypassing stack protector is possible {}, for this reason we simply disable stack protector for this PoC.

\vspace{.04in}
\noindent{\bf PoC Attack: Arbitrary Code Execution}
The victim program has a class hierarchy similar to the running example, with additional classes A1 and A2 which are proper base classes of A. In the main function, an object of class B is created. The main function has a buffer overflow vulnerability on line 22. On line 23 callBaseFunc() is called on B’s object which in turn calls function geta() defined in A (B's primary base). Instead of this intended call, the PoC attack diverts control to function execShell() defined in A2 (B's secondary base). Note that A's subobject in B is located at offset 0x10 from the top of B's object, while A2's subobject is located at offset 0x20.

First, we identified the address  of the construction VTable of B-in-D which has a vbaseoffset of 0x20. Next, we exploited the buffer overflow vulnerability to corrupt the address point of object b. We overflow buf2 (line 22) into b such that the vptr of B (which has a vbaseoffset of 0x10) is overwritten with that of the construction VTable of B-in-D. Recall that Marx will allow this since it does not differentiate between a regular and a construction VTable. We show the disassembly of function B::callBaseFunc() in Listing~\ref{callBaseFunc}. On line 3 of Listing~\ref{callBaseFunc}, the vbaseoffset is retrieved to locate A’s subject in B. Because of the buffer overflow, a vbaseoffset of 0x20 is retrieved thereby locating the subobject of A2 instead. As a result, line 7 locates the second virtual function in the VTable of A2 (execShell) and executes it. \ignore{This source code and the binary for this PoC is available at github...}

\lstset{language=C++}
\begin{lstlisting}[aboveskip=-\baselineskip, belowskip=-1.8\baselineskip, caption={Vulnerable program}, label={attack}, float,floatplacement=H]
1.  class A1{
virtual int geta1(){...}
2.  };
3.  class A2{ ...
4.   virtual int geta2(){...}
5.   virtual void execShell(){system("/bin/sh");}
6.  };
7.  class A: public A1, public A2{ ...
8.   virtual int geta(){...}
9.  };
10. class B: public virtual A{ ...
11.  virtual int callBaseFunc(){return geta();}
12. };
13. class C: public virtual A{ ...
14. };
15. class D: public B, public C{ ...
16. };
17. int main(){ ...
18.  B b;
19.  char buf2[10];
20.  char buf1[20];
21.  scanf("%20s", buf1);
22.  strcpy(buf2, buf1); //overflow buf2
23.  b.callBaseFunc();
24.  return 0;
25. }
\end{lstlisting}

\lstset{language=[x86masm]Assembler}
\begin{lstlisting}[belowskip=-1.8\baselineskip, caption={Disassembly of B::callBaseFunc()}, label={callBaseFunc}, float,floatplacement=H]
...
1. mov eax [b_obj_addr] ;get vptr of B
;Marx-like check: is_vptr_valid(eax) -> True
2. sub eax, 0xC ;locate vbase-offset field in VTable of B
;get vbaseoffset, 0x20 instead of 0x10 after overflow
3. mov eax, [eax] 
4. ...
5. add eax, b_obj_addr ;reach subobject A2 instead of A
6. mov eax, [eax] ;get vptr of A2 instead of vptr of A
;get second func in VTable, execShell() instead of geta()
7. add eax, 4 
8. ...
9. call eax
\end{lstlisting}
\ignore{
	\subsubsection{Data Corruption}
	While virtual inheritance can be exploited to perform limited arbitrary code execution (indirect call target must be in the allowable set), it can be exploited to achieve data corruption at any location (within or outside the object being used) in the program's memory accessible by a vbase-offset. Any access to a member of a virtual base from a derived object can be diverted to access any location in memory accessible at an offset equal to a vbase-offset in a construction VTable.\ignore{ corresponding to the derived class.}}

\subsection{Attack Surface Analysis}\label{attack_surface}
\begin{figure}
	\tikzset{every mark/.append style={scale=1.5}}
	\begin{tikzpicture}[scale=0.8]
	\begin{axis}[scatter/classes={
		mysqld={mark=x,MidnightBlue},
		mysqlbinlog={mark=x,Maroon},
		mysqlpump={mark=x,OliveGreen},
		libstdcm={mark=x,YellowOrange},
		libcmism={mark=x,WildStrawberry},
		darkicem={mark=x,Violet},
		dealm={mark=x,Thistle},
		ragelm={mark=x,draw=Black}},
	xtick=data,
	xticklabels={mysqld(48), mysqlbinlog(14), mysqlpump(4), libstdc++(22), libcmis(14), darkice(18), dealII(6), ragel(8)},
	x tick label style={rotate=45,anchor=east}]
	\addplot[scatter,only marks, scatter src=explicit symbolic] coordinates {
		(1, 8) [mysqld]
		(1, 16) [mysqld]
		(1, 24) [mysqld]
		(1, 40) [mysqld]
		(1, 48) [mysqld]
		(1, 56) [mysqld]
		(1, 64) [mysqld]
		(1, 72) [mysqld]
		(1, 80) [mysqld]
		(1, 112) [mysqld]
		(1, 120) [mysqld]
		(1, 128) [mysqld]
		(1, 240) [mysqld]
		(1, 248) [mysqld]
		(1, 256) [mysqld]
		(1, 264) [mysqld]
		(1, 272) [mysqld]
		(1, 280) [mysqld]
		(1, 288) [mysqld]
		(1, 296) [mysqld]
		(1, 304) [mysqld]
		(1, 368) [mysqld]
		(1, 608) [mysqld]
		(1, 888) [mysqld]
		
		(1, -8) [mysqld]
		(1, -16) [mysqld]
		(1, -24) [mysqld]
		(1, -40) [mysqld]
		(1, -48) [mysqld]
		(1, -56) [mysqld]
		(1, -64) [mysqld]
		(1, -72) [mysqld]
		(1, -80) [mysqld]
		(1, -112) [mysqld]
		(1, -120) [mysqld]
		(1, -128) [mysqld]
		(1, -240) [mysqld]
		(1, -248) [mysqld]
		(1, -256) [mysqld]
		(1, -264) [mysqld]
		(1, -272) [mysqld]
		(1, -280) [mysqld]
		(1, -288) [mysqld]
		(1, -296) [mysqld]
		(1, -304) [mysqld]
		(1, -368) [mysqld]
		(1, -608) [mysqld]
		(1, -888) [mysqld]
		
		(2, 288) [mysqlbinlog]
		(2, 16) [mysqlbinlog]
		(2, 8) [mysqlbinlog]
		(2, 112) [mysqlbinlog]
		(2, 144) [mysqlbinlog]
		(2, 24) [mysqlbinlog]
		(2, 120) [mysqlbinlog]
		
		(2, -288) [mysqlbinlog]
		(2, -16) [mysqlbinlog]
		(2, -8) [mysqlbinlog]
		(2, -112) [mysqlbinlog]
		(2, -144) [mysqlbinlog]
		(2, -24) [mysqlbinlog]
		(2, -120) [mysqlbinlog]
		
		(3, 8) [mysqlpump]
		(3, 38) [mysqlpump]
		
		(3, -8) [mysqlpump]
		(3, -38) [mysqlpump]
		
		(4, 96) [libstdcm]
		(4, 88) [libstdcm]
		(4, 248) [libstdcm]
		(4, 112) [libstdcm]
		(4, 8) [libstdcm]
		(4, 128) [libstdcm]
		(4, 256) [libstdcm]
		(4, 104) [libstdcm]
		(4, 264) [libstdcm]
		(4, 24) [libstdcm]
		(4, 120) [libstdcm]
		
		(4, -96) [libstdcm]
		(4, -88) [libstdcm]
		(4, -248) [libstdcm]
		(4, -112) [libstdcm]
		(4, -8) [libstdcm]
		(4, -128) [libstdcm]
		(4, -256) [libstdcm]
		(4, -104) [libstdcm]
		(4, -264) [libstdcm]
		(4, -24) [libstdcm]
		(4, -120) [libstdcm]
		
		(5, 64) [libcmism]
		(5, 32) [libcmism]
		(5, 8) [libcmism]
		(5, 16) [libcmism]
		(5, 72) [libcmism] 
		(5, 24) [libcmism]
		(5, 40) [libcmism] 
		
		(5, -64) [libcmism]
		(5, -32) [libcmism]
		(5, -8) [libcmism]
		(5, -16) [libcmism]
		(5, -72) [libcmism] 
		(5, -24) [libcmism]
		(5, -40) [libcmism] 
		
		(6, 32) [darkicem] 
		(6, 64) [darkicem] 
		(6, 56) [darkicem] 
		(6, 40) [darkicem] 
		(6, 96) [darkicem] 
		(6, 80) [darkicem] 
		(6, 104) [darkicem] 
		(6, 112) [darkicem] 
		(6, 24) [darkicem] 
		
		(6, -32) [darkicem] 
		(6, -64) [darkicem] 
		(6, -56) [darkicem] 
		(6, -40) [darkicem] 
		(6, -96) [darkicem] 
		(6, -80) [darkicem] 
		(6, -104) [darkicem] 
		(6, -112) [darkicem] 
		(6, -24) [darkicem] 
		
		(7, 176) [dealm] 
		(7, 168) [dealm] 
		(7, 336) [dealm] 
		
		(7, -176) [dealm] 
		(7, -168) [dealm] 
		(7, -336) [dealm] 
		
		(8, 8) [ragelm]
		(8, 136) [ragelm]
		(8, 16) [ragelm]
		(8, 24) [ragelm]
		
		(8, -8) [ragelm]
		(8, -136) [ragelm]
		(8, -16) [ragelm] 
		(8, -24) [ragelm] 

	};
	\end{axis}
	
	\end{tikzpicture}
	\caption{Distribution of vbase-offset and offset-to-top from construction VTables. The number in parentheses next to the binary name is the exact number of unique vbase-offset and offset-to-top values for that binary.}\label{cvtable-distr}
\end{figure}
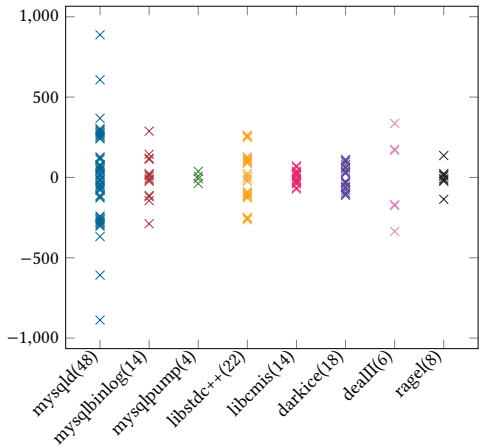

The attack surface that manifests due to the presence of virtual inheritance directly relates to the number of construction VTables. That is, the number of offsets that can be exploited increases with the number of construction VTables present in the binary, especially if they contain sufficient unique offsets. 
Unique vbase-offset and offset-to-top values for representative realworld programs in our study is presented in Figure~\ref{cvtable-distr}. 
Results indicate that it is not uncommon for offset values to be in multiple hundreds, which in turn indicates potential for an attacker to perform memory corruption attacks multiple hundred bytes from an intended access. 

In general, attack surface increases with the number of construction VTables, which in turn increases \textit{exponentially} with depth of inheritance. 
Table~\ref{vtable_explosion} presents the number of construction VTables for the running example (row 1) and increasing depths (up to {\em n}).
Total number of construction VTables at depth {\it n} is:
\begin{equation}
\sum (n+1) -1 = \frac{(n+1)(n+2)}{2}-1 \implies O(n^2)
\end{equation}
Since each derived class in the virtual inheritance tree may have several varying object layouts, there is a high probability that an attacker will find sufficient unique offsets needed to carry out an attack. For instance, in mysqld, there are 6 unique virtual bases in the entire inheritance tree, however, we found 24 unique vbase-offsets and 24 unique offset-to-top values. Furthermore, in program's that use libraries such as libstdc++, inheriting from virtual bases (e.g., Stream class) will introduce additional construction VTables in a program's memory. 

\begin{table}[ht]
	\centering
	\normalsize
	\caption{Table showing how the number of construction VTables increase with inheritance depth}
	\label{vtable_explosion}
	\scalebox{0.9}{%
		\begin{tabular}{ll}
			\toprule[0.4ex]
			\textbf{Inheritance tree} & \textbf{No of construction VTables} \\ \midrule
			As in Listing~\ref{virtualinh} (depth 1) & 2 \\ 
			E inherits from D (depth 2) & 2+3 = 5 \\ 
			F inherits from E (depth 3) & 2+3+4 = 9 \\ 
			X inherits from Y (depth n) & 2+3+4+...+n+1 = $\sum (n+1) -1$ \\ 
			\bottomrule[0.4ex]
		\end{tabular}
	}
\end{table}

\section{Solution Overview}\label{sec:overview}
\begin{figure*}
	\centering
	\includegraphics[scale=.5]{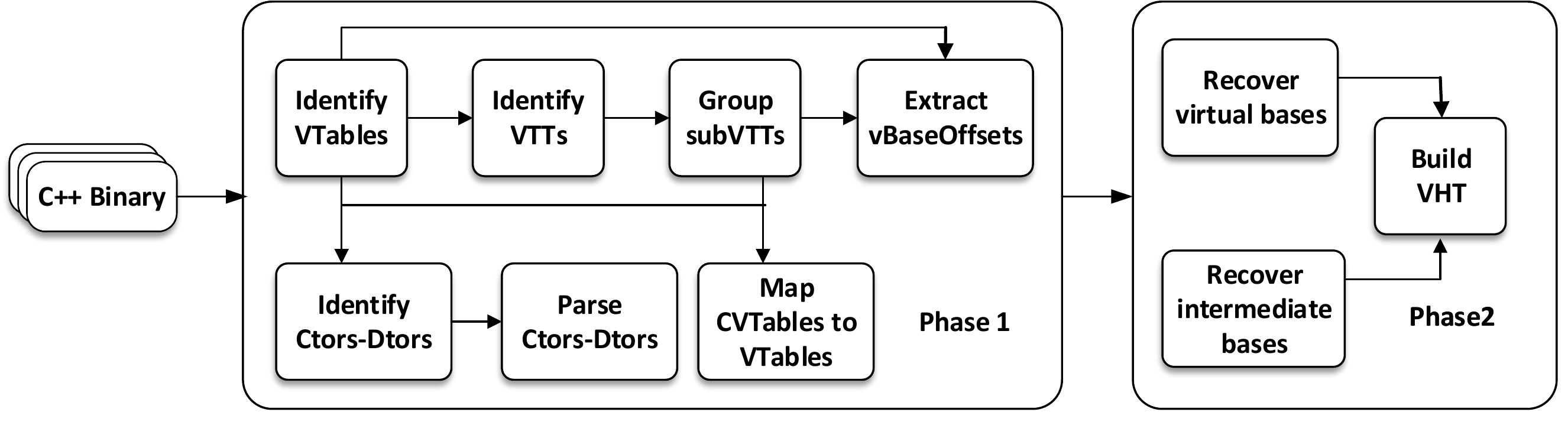}
	\caption{Overview of \codename}\label{fig:overview}
\end{figure*}

An overview of our solution \codename\ is presented in Figure ~\ref{fig:overview}. \codename\ tackles multiple challenges posed by virtual inheritance.

\subsection{Challenges in Recovering Virtual Inheritance}
\paragraph{Presence of Optional Fields} Unlike single and multiple inheritance where mandatory fields such as offset-to-top and type-info (i.e., RTTI) provide a reliable signature, virtual inheritance introduces \textit{one or more} optional fields vcall-offsets and vbase-offsets which makes analysis difficult.
These fields pose 2 main challenges. First, without knowing how many optional entries are present, identifying the boundaries of VTables is hard. Second, because one or more entries of vcall-offsets could be laid immediately preceding one or more entries of vbase-offsets, it is hard to demarcate the end of vcall-offsets and beginning of vbase-offsets. 

\ignore{
	fields in a VTable are fixed, virtual inheritance introduces optional fields which makes identifying the boundary of a VTable difficult. Specifically, in single and multiple inheritance, there is exactly one field each for offset-to-top and type-info (i.e., RTTI). However, with virtual inheritance, in addition to offset-to-top and type-info that are mandatory fields, there could be \textit{one or more} vbase-offsets (depending on the number of virtual bases) and \textit{one or more} vcall-offsets (depending on the number of virtual functions defined by a virtual base). Precise analysis is necessary to delineate between the various fields.
}

\paragraph{Construction VTables vs Regular VTables} The layout of a construction VTable and a regular VTable are exactly the same.
However, they have different purposes. Therefore a trivial signature-based approach is insufficient to distinguish between the two. \ignore{A successful solution must employ cross-semantics analysis (i.e., combine information available in class layout/constructors with VTables) in order to distinguish between regular and construction VTables. }
Moreover, a class may contain multiple construction VTables depending on the depth of inheritance between the virtual base and the derived class. In order to build a clear and accurate class hierarchy tree, we must be able to group all the construction VTables of a class and associate them with the complete object VTable of that class.

\ignore{
	\paragraph{Grouping Construction VTables Belonging to a Class}
	A derived class has at least as many construction VTables as the number of its intermediate bases (detailed analysis in Section~\ref{attack_surface}). In order to build a clear and accurate class hierarchy tree, we must be able to group all the construction VTables of a class and associate them with the complete object VTable of that class. This will ensure that only the complete object VTable is represented in the class hierarchy tree.
}

\paragraph{Differentiating Virtual and Non-virtual Bases} 
When a class derives from both virtual and non-virtual bases, its object and VTable contain subobjects and subVTables that correspond to both base classes. In order to correctly reconstruct the virtual inheritance tree, there is the need to identify non-virtual bases and filter them out.

\subsection{High-Level Approach}
The Figure~\ref{fig:overview} shows the overview of \codename. It incorporates analysis passes that tackle the challenges described above.

\paragraph{Discerning Relationship Between Mandatory and Optional Fields}
There is simply not enough information in one VTable alone to demarcate different optional and mandatory fields. Therefore, our analysis combines information in multiple VTables.  For instance, the offset-to-top value from a secondary VTable corresponding to a virtual base is equal in magnitude to the vbase-offset in the derived object's primary VTable. Such a correlation provides a strong confirmatory test to filter vbase offsets.  Additionally, when the types are statically known during compilation, a vbase-offset is applied by the compiler during computation of vbase object's address from a derived object's address.  By cross referencing potential offsets in the VTable with offseting code emitted by the compiler, it is possible to identify optional offsets with a high level of confidence. 

\ignore{
	Starting from the address-point, we first identify the mandatory fields in the VTable. Then, we correlate the offset information from the  offset information with the optional fields to correctly recover virtual bases. That is, the vbase-offset from a derived object to a virtual base sub-object is equal in magnitude (but opposite in sign) to the offset-to-top from virtual base sub-object to the derived object.
}

\ignore{
	Apart from the virtual functions and the type-info (RTTI) fields in a VTable, every other field represents offset from or to subobjects within an object. All these fields, mandatory and optional have certain relationships. For instance, the offset-to-top value of a virtual base subobject is equal to the negation of the vbase-offset from the derived subobject to the virtual base subobject. Starting from the address-point, \codename\ first identifies the mandatory fields in the VTable. Then, it correlates the mandatory offset information with the optional fields to correctly recover virtual bases. That is, the vbase-offset from a derived object to a virtual base sub-object is equal in magnitude (but opposite in sign) to the offset-to-top from virtual base sub-object to the derived object. 
}

\paragraph{Identifying Construction VTables}
\codename\ incorporates VTT analysis in order to identify construction VTables. The VTT is a key signifier of virtual inheritance in a binary, however, it is also crucial for differentiating construction VTables from regular VTables. Per ABI mandate, the first entry in a VTT always points to a regular VTable, and every other entry points to a construction VTable. \codename\ first identifies VTTs in the binary and then isolates the regular and construction VTables from the first and remaining entries in a VTT. 

\paragraph{Grouping Construction VTables of a Class}
The virtual function fields in all construction VTables of a class and its regular VTable are exactly the same. \codename\ takes advantage of this similarity to identify and group all the construction VTables belonging to a class.

\paragraph{Identifying Virtual and Non-virtual bases}
Only virtual bases have associated optional fields such as vbase- and vcall-offset. Since \codename\ has already recovered the optional fields in an earlier step, it can filter out non-virtual bases from the inheritance tree.

\section{VirtAnalyzer}\label{sec:design}
The \codename\ consists of two phases, each of which consists of multiple sub-phases. \ignore{To recover virtual inheritance from a binary, we first identity VTables, then we identify the VTTs it contains, then we get subVTTs from the VTTs and finally, we extract virtual base offsets form VTables.} We show the data used in each sub-phase in Table~\ref{stepsndata}. We explain the phases and their sub-phases below.

	\begin{table}[ht]
		\centering
		\normalsize
		\caption{Table showing steps in recovering virtual inheritance and the data involved in each step}
		\label{stepsndata}
		\rowcolors{1}{white}{gray!30}
		\scalebox{0.8}{%
			\begin{tabular}{lc}
				\toprule[0.4ex]
				Step & Data involved \\ \midrule
				Identifying VTables & VTables \\
				Identifying VTTs & VTables, VTTs \\
				Grouping subVTTs & VTTs, subVTTs \\
				\specialcell{Extracting virtual \\base offsets} & VTables, subVTTs \\
				\specialcell{Mapping constructor VTables\\ to regular VTables} & \specialcell{Construction VTables,\\ VTables, VTTs,\\ subVTTs} \\
				\specialcell{Identifying constructors\\ destructors} & Ctors, Dtors, VTables \\
				Parsing ctors and dtors & Ctors, Dtors \\
				Recovering virtual bases & \specialcell{VTables, Ctors, Dtors,\\ vbaseoffset} \\
				Recovering intermediate bases & subVTTs, Ctors, Dtors \\ \bottomrule[0.4ex]
			\end{tabular}
		}
\end{table}

\ignore{
	\begin{algorithm}[ht]
		\caption{RecoverVirtualInheritance.} \label{alg:virtinhrec}
		\begin{algorithmic}[1]
			\footnotesize
			\Procedure{RecoverVirtualInheritance}{}
			\State $VTables \gets getVTables()$
			\State $VTTs \gets scanRodataForVTTs(VTables)$
			\State $subVTTs \gets breakVTTsIntoSubs(VTTs, VTables)$
			\State $vbaseoffs \gets recAndVerifyVbaseoffs(subVTTs, VTables)$
			\State $ctorsAnddtors\gets identifyCtorDtor()$
			\State $parsedInstns\gets parseInstns(ctorsAnddtor)$
			\State $mergeVTables(VTables, cVTables)$
			\State $virtualBases\gets getVirtuals(parsedInstns, vbaseoffs, ctorsAnddtors)$
			\State $intermBases\gets getInterm(parsedInstns, ctorsAnddtors)$
			\State $finalizeVirtInh(virtualBases, intermBases)$
			
			\EndProcedure
			
		\end{algorithmic}
\end{algorithm}}

\subsection{Phase 1: Extracting Metadata from the Binary}\label{subsec:extraction}
In this phase, we recover information in the binary which indicate virtual inheritance. The sub-phases here include identifying certain structures, values and functions, such as VTables, VTTs, subVTTs, vbaseoffsets, constructors and destructors.

\vspace{.04in}
\noindent{\bf Identifying VTables}
We identified VTables by implementing algorithms presented in vfGuard~\cite{prakash:2015:vfguard} and \codenametwo~\cite{declassifier}. Usually, VTables are referenced from the text section using immediate values that point to the read-only section. \ignore{First, we scan through the text section to recover all immediate values pointing to the data section.}However, we found some libraries with immediate values in the text section pointing to the got section, which then point to the VTables. Such cases too were handled in our analysis. We recover immediate values and examine them to see if they are vptrs. An immediate value is a vptr if the output from dereferencing:
\begin{itemize}
	\item  vptr points to a function start address or the pure virtual function and
	\item (vptr - DWORD\_SIZE) is either zero, or points to the data section (the typeinfo) and
	\item (vptr - DWORD\_SIZE*2) is zero (for a primary VTable) or a negative value (for a secondary VTable)
\end{itemize}
We identify valid vptrs (for both primary and secondary VTables) and then group primary vptrs with their corresponding secondary vptrs to obtain the complete object VTable of each class. This was done by implementing the VTable grouping algorithm introduced in \codenametwo~\cite{declassifier}. \ignore{VTable grouping is based on the idea that VTables of a class are laid out contiguously, starting with the primary VTable and followed by all its secondary VTables. Therefore, the identified vptrs are sorted in order of increasing addresses. The first address in the list should belong to a primary VTable, it is grouped with all secondary vptrs following it, the next primary vptr encountered belongs to another class.} This set of VTables also includes construction VTables.

\vspace{.04in}
\noindent{\bf Identifying VTTs}
VTTs, like VTables also reside in the read-only section of the binary. They are the only structures whose entries are pointers to VTables. We identify VTTs by first identifying structures that contain at least two entries of pointers to known VTables. 

Unlike VTables whose offset-to-top field or typeinfo field can be used to separate two VTables which are laid out contiguously, VTTs only contain pointers, as a result, it is tricky to identify the boundaries between VTTs if they are laid out contiguously. In most cases, the VTables pointed to by a VTT are laid out immediately after the VTT which makes it easy to know where a VTT ends. However, we found a few cases where VTTs of different classes are laid out contiguously. In these cases there is the possibility of wrongly grouping those multiple VTTs as one VTT. To address this, we take advantage of how VTT entries are ordered. 

Notice from Figure~\ref{fig:vtt} that the VTT of D starts with the vptr for D's VTable, followed by all the construction VTables (this means the second entry belongs to a construction VTable), and ends with D's secondary VTables. We also noticed that construction VTables are laid out after their Derived VTables. We first store the values of the first and second entries of a VTT and continue down the VTT. We say that the previous entry found is the last entry in a VTT if the value of the current entry is either less than the first entry (the last entry should be greater than the first since it belongs to a secondary VTable) or greater than the second entry (the last entry should be less then the second since construction VTables are laid out after the derived VTable). \ignore{We do this by scanning every DWord address in the rodata section, and then verifying if it points to a valid VTable. Only VTT entries are pointers in rodata section which point to VTables. Once we find such address, we scan below it to recover all other entries associated to it (they belong to the same VTT). We stop scanning for that VTT when we encounter a value that is not a pointer to a VTable.} Algorithm~\ref{alg:vtt} shows how VTTs are identified.

\begin{algorithm}[ht]
	\caption{IdentifyAVTT.} \label{alg:vtt}
	\begin{algorithmic}[1]
		\footnotesize
		\Procedure{IdentifyAVTT}{$addr, nextVTTIndex$}
		\State {$vptr \gets getvptrAtAddr(addr)$}
		\If {$isInSegment(vptr, "data")$}
		\If {$isAValidVTable(vptr)$}
		\State {$newVTT \gets \emptyset$}
		\State {$newVTT.append(addr)$}
		\State {$nextEntry \gets getNextVTTEntry(addr)$}
		\While {$nextEntry != -1$}
		\State {$nextVptr \gets getvptrAtAddr(nextEntry)$}
		\If {$nextVptr < vptr$}
		\State {\bf $break$}
		\EndIf
		\If {$isAValidVTable(nextVptr)$}
		\State {$newVTT.append(nextEntry)$}
		\Else
		\State {\bf $break$}
		\EndIf 
		\State {$nextEntry \gets getNextVTTEntry(nextEntry)$}
		\EndWhile
		\If {$len(newVTT) > 1$}
		\State {$VTTs[nextVTTIndex] \gets newVTT$}
		\State {$nextVTTIndex++$}
		\EndIf
		\EndIf
		\EndIf
		
		\Return $nextVTTIndex$
		
		\EndProcedure
		
	\end{algorithmic}
\end{algorithm}

\vspace{.04in}
\noindent{\bf Grouping subVTTs}
A VTT has pointers to a VTable and one or more construction VTables. We refer to a group of pointers to each of these complete object VTables as subVTT. The first subVTT in a VTT points to the regular VTable of the derived class, while the other subVTTs point to its construction VTables. We differentiate regular VTables from construction VTables by looking at their subVTT position in the VTT. To identify subVTTs, we start scanning a VTT from the beginning. Anytime we find a pointer to a primary VTable, we create a new subVTT, and all secondary VTable found (before the next primary VTable) are grouped together. We are able to prevent grouping the secondary VTables of the derived class with the last construction VTable because the secondary VTables' addresses will be less than those of the construction VTables. Algorithm~\ref{alg:subvtt} presents the specific steps in grouping subVTTs.

\begin{algorithm}[ht]
	\caption{GroupSubVTTs.}\label{alg:subvtt}
	\begin{algorithmic}[1]
		\footnotesize
		\Procedure{GroupSubVTTs}{aVTT, VTables}
		\State $ordered\_vptr \gets \emptyset$
		\For {{\bf each} $addr$ {\bf in} $aVTT$}
		\State $ordered\_vptr.append(addr)$
		\EndFor
		\State $ordered\_vptr.sort()$
		\State $k = -1$
		\For {{\bf each} $addr$ {\bf in} $ordered\_vptr$}
		\State $ott \gets VTables[addr]['offsetToTop']$
		\If {$ott == 0$}
		\State $k = addr$
		\State $SubVTTs[k] \gets \emptyset$
		\State $SubVTTs[k].append(k)$
		\Else
		\If {$k == -1$}
		\State $continue$
		\EndIf
		\State $SubVTTs[k].append(addr)$
		\EndIf
		\EndFor
		\EndProcedure
	\end{algorithmic}
\end{algorithm}

\vspace{.04in}
\noindent{\bf Extracting Virtual Base Offsets}
There is the possibility of recovering false VTables and this will invariably result in recovering false VTTs. A group of pointers may point to recovered false VTables, we will wrongly identify this as a VTT. However, by recovering and verifying the vbase-offset they contain, we will realize that they are invalid. The vbase-offset a class is also instrumental in recovering its virtual bases. Algorithm~\ref{alg:vbo} shows the steps used in this sub-phase. 

The vbase-offset is one of the optional fields a VTable may contain depending on whether it is in a virtual inheritance tree or not. The number of vbase-offsets a derived class has is equal to the number of its virtual bases. \ignore{A vbase-offset is equal to the negation of the offset-to-top of one of derived class' virtual bases (the offset to top will be present in the secondary VTable corresponding to that base).} Since a VTable can contain multiple vbase-offsets, we  recover the first vbase-offset by deferencing the third Dword from the vptr (upward direction) and then we go up the VTable one Dword at a time to recover the rest. As we recover each vbase-offset, we verify it. This is done by comparing it with the offset-to-top values in the secondary VTables from the first to the last. If the current vbase-offset is not equal to the negative value of the offset-to-top of current secondary VTable, we assume the base corresponding to that VTable is not a virtual base, then we move to the next secondary VTable. We stop scanning for vbase-offset when the number of Dwords above the vptr checked is equal to the number of secondary VTables. We must find at least one matching vbase-offset and offset-to-top to conclude that the subVTT is valid. If we do not find any valid subVTT in a VTT, we discard the VTT.

\begin{algorithm}[ht]
	\caption{ExtractVBaseOffsets.} \label{alg:vbo}
	\begin{algorithmic}[1]
		\footnotesize
		\Procedure{ExtractVBaseOffsets}{aSubVTT, VTables}
		\State $vptr \gets getPryVptr(aSubVTT)$
		\State $curLoc \gets vptr - (DWORD\_SIZE * 3)$
		\State $vBaseOffs[vptr] \gets \emptyset$
		\For {{\bf each} $i$ {\bf in} $aSubVTT$}
		\State $ott \gets VTables[i]['offsetToTop']$
		\State $vbo \gets Dword(curLoc)$
		\If {$ott == neg(vbo)$}
		\State $vBaseOffs[vptr].append(vbo)$
		\State $curLoc \gets curLoc - DWORD\_SIZE$
		\EndIf
		\EndFor
		\EndProcedure
	\end{algorithmic}
\end{algorithm}

\vspace{.04in}
\noindent{\bf Mapping Construction VTables to Regular VTables}\label{map_vts}
As stated in section~\ref{vtt}, only the first subVTT in a VTT belong to the VTable of the derived class, the other subVTTs belong to construction VTables of IntermediateBase-in-Derived. We use this information to differentiate a construction VTable from a regular VTable. In this sub-phase, we identify the corresponding complete object VTable of every construction VTable. One or more construction VTables can map to one complete object VTable. For instance, in the running example, we map the construction VTable of B-in-D to the VTable of B. The mapping is constructed based on the observation that the function pointers in a complete object VTable and its corresponding construction VTables are exactly the same. To construct the mapping, we sum up the function pointers in each VTable and use that as the key in a dictionary. A regular VTable and its corresponding construction VTables will have the same key. This mapping is needed while building the virtual inheritance tree.

\vspace{.04in}
\noindent{\bf Identifying Constructors and Destructors}\label{iden_cdtor}
Constructors and destructors are where vptr initializations take place, for both object construction and destruction. We identify constructors and destructors, using the same approach employed by existing solutions. A function is said to be a constructor or destructor if it initializes an immediate value which points to a known VTable. \ignore{we scan the .text section of the binary for functions containing instructions that initialize vptrs.}

\vspace{.04in}
\noindent{\bf Parsing Constructors and Destructors}
We analyze each instruction within constructors and destructors to keep track of how offset in an object and VTT addresses are propagated from one register or memory location to another. We use this information to identify and retrieve the hidden arguments (\texttt{this} pointer and/or subVTT address) passed to regular constructors and destructors as well as the special constructors and destructors. We also keep track of call instructions which will be used in phase 2 of the analysis to identify virtual and intermediate bases.

\subsection{Phase 2: Recovering Bases}\label{subsec:recovery}
\ignore{We explain steps to recover virtual and intermediate bases using Figure~\ref{fig:cases}. Diagram 1 has only direct virtual bases, diagram 2 has 2 direct virtual base, while diagram 3 has one indirect virtual base. The non-virtual base in diagram 2 is recovered by the well studied class hierarchy recovery approach. We discuss the approach to recover the virtual bases in diagrams 1 and 3, as well as the approach to recover the intermediate bases B and C in diagram 3. \ignore{We provide the disassembly for the constructor of D in diagram 1.} Diagram 3 is the same as the running example whose disassembly is given in Listing~\ref{l3}.}
Virtual inheritance can result from either direct (i.e., a direct base of a derived class is virtually inherited) or indirect (i.e., an intermediate base class of a derived class virtually inherits from its base). Specific details can be found in 2.5.3 of the ABI~\cite{ItaniumABI}. Further, it is possible that a derived class has a mix of virtual and non-virtual bases.  

\ignore{
	\begin{lstlisting}[caption={Disassembly of the ctor of D in Figure 1}, label={l1}]
	...
	1.  mov [rbp+var_8], rdi
	2.  mov rax, [rbp+var_8]
	3.  add rax, 10h
	4.  mov rdi, rax; this, at offset 10h
	5.  call _ZN1AC2Ev; A::A(void)
	...
	6.  mov rax, [rbp+var_8]
	7.  add rax, 20h
	8.  mov rdi, rax; this, at offset 20h
	9.  call _ZN1BC2Ev; B::B(void)
	...
	10. mov rax, [rbp+var_8]
	11. add rax, 30h
	11. mov rdi, rax; this, at offset 0
	12. call _ZN1CC2Ev; C::C(void)
	...
	\end{lstlisting}}

\ignore{
	\begin{lstlisting}[caption={Disassembly of the ctor of D in diagram 2}, label={l2}]
	...
	1.  mov [rbp+var_8], rdi
	2.  mov rax, [rbp+var_8]
	3.  mov rdi, rax; this, at offset 0
	4.  call _ZN1AC2Ev; A::A(void) primary base class
	...
	5.  mov rax, [rbp+var_8]
	6.  add rax, 10h
	7.  mov rdi, rax; this, at offset 10h
	8.  call _ZN1BC2Ev; B::B(void)
	...
	9.  mov rax, [rbp+var_8]
	10. add rax, 20h
	11. mov rdi, rax; this, at offset 20h
	12. call _ZN1CC2Ev; C::C(void)
	...
	\end{lstlisting}
}

\vspace{.04in}
\noindent{\bf Recovering Virtual Bases}\label{rec_virtual_base}
\ignore{\ayomide{maybe I should name this differently, because it recovers direct or topmost virtual bases}}
The constructor of a derived class directly calls the constructors of its virtual bases, be it direct or indirect, along with the constructors of its direct non-virtual bases. Before calling the constructor of a virtual base, an offset equal to the vbase-offset corresponding to that base is added to the {\tt this} pointer. For example, in Listing~\ref{l1}, lines 3, 7 and 11, offsets of 0x10, 0x20 and 0x30 respectively (vbase-offset of A-in-D,B-in-D and C-in-D) are added to the {\tt this} pointer before the call instructions on lines 5, 9 and 12 respectively. Similarly, in Listing~\ref{l3}, line 3, offset of 0x20 (vbase-offset of A-in-D) is added to the {\tt this} pointer. \ignore{Concrete examples can be found in appendix~\ref{vbase_offset_example}.}

\begin{lstlisting}[caption={Disassembly of the ctor of D in Figure 1}, label={l1}, float,floatplacement=H]
...
1.  mov [rbp+var_8], rdi
2.  mov rax, [rbp+var_8]
3.  add rax, 10h
4.  mov rdi, rax; this, at offset 10h
5.  call _ZN1AC2Ev; A::A(void)
...
6.  mov rax, [rbp+var_8]
7.  add rax, 20h
8.  mov rdi, rax; this, at offset 20h
9.  call _ZN1BC2Ev; B::B(void)
...
10. mov rax, [rbp+var_8]
11. add rax, 30h
11. mov rdi, rax; this, at offset 0
12. call _ZN1CC2Ev; C::C(void)
...
\end{lstlisting}

In order to identify a virtual base, we scan the constructor of the derived base for calls to other constructors, we then analyze the offsets added to the {\tt this} pointer before the calls are made. If an offset equals any vbase-offset found in the derived class' primary VTable, we conclude that the constructor being called belongs to a virtual base of the derived class. Lastly, we retrieve the primary vptr corresponding to the identified virtual base \ignore{using the mapping created in subsection~\ref{iden_cdtor}} and then record it as a virtual base vptr. \ignore{We show the steps to recover virtual bases in Algorithm~\ref{alg:vbase}
	
	\begin{algorithm}[ht]
		\caption{RecoverVirtualBase.} \label{alg:vbase}
		\begin{algorithmic}[1]
			\footnotesize
			\Procedure{RecoverVirtualBases}{parsedInstns, vBaseOffs, ctors\_dtors, call\_instns}
			\For {{\bf each} $v$ {\bf in} $vBaseOffs$}
			\For {{\bf each} $cd$ {\bf in} $ctors\_dtors[v]$}
			\For {{\bf each} $(ea, callee)$ {\bf in} $call\_instns[cd]$}
			\State $objAddr \gets getAddrPassedTo(ea)$ 
			\For {{\bf each} $off$ {\bf in} $objAddr$}
			\If {$rdi$ {\bf in} $objAddr[off]$ \&\&  $off$ {\bf in} $vBaseOff[v]$}
			\State $vBase[v].append(getPryVptr(callee))$
			\EndIf
			\EndFor
			\EndFor
			\EndFor
			\EndFor
			\EndProcedure
		\end{algorithmic}
\end{algorithm}}

\vspace{.04in}
\noindent{\bf Recovering Intermediate Bases}\label{rec_intermediate}
A special constructor is used to construct intermediate bases, virtual or non-virtual. This special constructor has three major distinctions from a regular constructor. It takes two default argument, {\tt this} pointer and subVTT address (lines 7 and 13 of Listing~\ref{l3}), unlike the regular constructors whose only default argument is the {\tt this} pointer. The subVTT address contains pointers to the construction VTable of the intermediate-in-derived. Second, it does not call the constructors of any of its virtual bases (they are called by the classes that derive from it). Third, it does not initialize its vptrs using immediate values, rather it accesses the subVTT it received as argument to get the vptrs needed for initialization.

The third distinction will make the well known method of identifying constructors and destructors to fail since there is no vptr initialization. The calls on lines 10 and 16 of Listing~\ref{l3} will be seen as just a function call. As a result, no relationship will be identified between B and D, and C and D. To address this problem, we make use of another information that the constructor of a derived class exposes. For every intermediate base, the derived class calls a special constructor which takes a subVTT address (an immediate value) as its second argument. Therefore, to recover intermediate bases, we scan constructors for subVTT addresses. All those addresses represent individual intermediate bases. Once we have all the subVTT addresses, we retrieve their corresponding VTable addresses from the map obtained in subsection~\ref{map_vts} and then record them as intermediate bases of the derived class. \ignore{Algorithm~\ref{alg:ibase} shows the step used in this sub-phase.
	
	\begin{algorithm}[ht]
		\caption{RecoverIntermediateBase.} \label{alg:ibase}
		\begin{algorithmic}[1]
			\footnotesize
			\Procedure{RecoverIntermBases}{parsedInstns, ctorsAnddtors, subVTTs}
			\State $ctors\_dtors \gets getCDtorThatWriteVTTs()$
			\For {{\bf each} $cd$ {\bf in} $ctors\_dtors$}
			\State $VTTaddrs \gets getVTTsWritten(cd)$
			\State $vptr \gets getPryVptr(cd)$
			\For {{\bf each} $vtt$ {\bf in} $VTTaddrs$}
			\If {$hasCorrVptr(vtt)$}
			\State $intermBase[vptr].append(getCorrVptr(vtt))$
			\EndIf
			\EndFor
			\EndFor
			\EndProcedure
		\end{algorithmic}
\end{algorithm}}

\vspace{.04in}
\noindent{\bf Building Virtual Inheritance Tree}
Once we recover all classes involved in virtual inheritance, including their virtual and intermediate bases, we merge the results to construct the virtual inheritance tree of the binary. In order to show how our solution integrates with existing class inheritance recovery tools, we also build the overall inheritance tree which includes single, multiple and virtual inheritance.

\subsection{Support for MSVC ABI}\label{msvc}
MSVC ABI’s implementation of structures used to implement virtual inheritance is slightly different from Itanium ABI’s implement. Unlike the Itanium ABI, vbase-offsets, offset-to-top, RTTI (if present) and virtual functions are together within the VTables, MSVC ABI separates offsets, vbase-offsets are placed in a table named the virtual base table (VB-Table) while RTTI and virtual functions are placed in the VTable. \ignore{\color{red}Using the running example in Listing~\ref{virtualinh}, we will explain the content of these structures and how they are used.
	
	The offset in a polymorphic object points to (in the case of multiple inheritance, one of) its VTable, the next offset points to its VB-Table. A VB-Table is created for every class (polymorphic or non-polymorphic) that directly or indirectly inherits from a virtual base. Classes B, C and D in the running example all have VB-Tables. B and C have one each and D has two, one for D’s subobject (shared with B) and the other for C’s subobject. The VB-Tables are written to the object during construction and destruction.}

We recover virtual inheritance from a binary that follows the MSVC ABI implementation by first recovering VTables, this process is the same for the Itanium ABI. Next, we recover VB-Tables (Algorithm~\ref{alg:vbtable}). VB-Tables seem more difficult to precisely recover unlike VTables, however, we noticed that the first entry in every VB-Table we found is a particular constant value. We used that constant value as a signature to recover them. We recover constructors and destructor next, by looking for functions which initialize vptrs (same for Itanium ABI). The first VB-Table of a given type contains offset(s) to the virtual base(s) from the top of the object. Therefore to identify virtual bases (Algorithm~\ref{alg:vbases_msvc}), we look for calls in constructors and destructors whose first argument is the address of the object plus an offset contained in the first VB-Table of the object type (similar to Itanium ABI). VTTs are not present in MSVC ABI binaries, therefore, intermediate bases are recovered differently. Since intermediate bases have direct or indirect virtual bases, they initialize VB-Table ptrs. We recover intermediate bases by looking for calls to constructors which initialize known VB-Table ptrs ((Algorithm~\ref{alg:vbases_msvc})). Note: MSVC ABI does not have construction VTables, however, the VB-Tables contain similar offset values as construction VTables. Therefore, virtual inheritance in a binary that adheres to MSVC ABI can be similarly exploited.
\begin{algorithm}[ht]
	\caption{GetVbtables.}\label{alg:vbtable}
	\begin{algorithmic}[1]
		\footnotesize
		\Procedure{GetVbtables}{}
		\State $imms \gets getImmediatesFrmText()$
		\For {{\bf each} $i$ {\bf in} $imms$}
		\If {$Dword(i) != getVbtableConstant()$}
		\State $continue$
		\EndIf
		\State $entries \gets \emptyset$
		\State $nLoc \gets i + DWORDSIZE$
		\While {$True$}
		\If {$Dword(nLoc) > 0 \&\& Dword(nLoc) < CAPOFFSET$}
		\State $entries.add(Dword(nLoc))$
		\State $nLoc \gets nLoc + DWORDSIZE$
		\Else
		\State $break$
		\EndIf
		\If {$len(entries) > 0$}
		\State $vbtables[i] \gets entries$
		\EndIf
		\EndWhile
		\EndFor
		
		\EndProcedure
	\end{algorithmic}
\end{algorithm}

\begin{algorithm}[ht]
	\caption{RecoverVBasesAndIBases.}\label{alg:vbases_msvc}
	\begin{algorithmic}[1]
		\footnotesize
		\Procedure{RecoverVBases}{ctorsDtors, vbtables}
		\For {{\bf each} $cd$ {\bf in} $ctorsDtors$}
		\State $vbptr \gets getFirstVbaseptr(cd)$
		\State $vptr \gets getAssVtable(cd)$
		\State $callInstns \gets getCallInstns(cd)$
		\For {{\bf each} $(addr, targ)$ {\bf in} $callInstns$}
		\State $addedOffset \gets getAddedOffset(addr)$
		\State $vptr\_t \gets getAssVtable(targ)$
		\If {$addedOff$ {\bf in} $vbtables[vbptr]$}
		\State $VirtualBases[vptr].add(vptr\_t)$
		\EndIf
		\If {$initVbptr(targ)$}
		\State $IntermBases[vptr].add(vptr\_t)$
		\EndIf
		\EndFor
		\EndFor
		
		\EndProcedure
	\end{algorithmic}
\end{algorithm}

\section{Implementation and Evaluation}\label{eval}
We developed \codename\ as an IDA Python plugin that builds on top of \codenametwo, an existing class hierarchy inference engine that infers single and multiple inheritances. 
\ignore{It is based on \codenametwo, an existing class hierarchy inference engine. \codename\ uses all the techniques used by \codenametwo\ to identify VTables, constructors and destructors, and single and multiple inheritance.} 
\codename\ can infer virtual inheritance from binaries that adhere to both Itanium and MSVC ABIs irrespective of the compiler used to compile the program (gcc, clang or visual studio). \ignore{Also, our current implementation supports 64-bit architecture. The only modification needed to support 32-bit architecture will be to change the size of a Dword (from 8 bytes to 4 bytes)}.

\ignore{
	\codename\ recovers all data structures used by scanning sections of the binary. VTables, VTTs, subVTTTs and vbase-offsets are recovered from the data section while constructors and destructors are recovered from the text section.
}
We aim to answer the following questions in our evaluation:
\begin{itemize}
	\item How accurately can virtual inheritance tree be recovered from a stripped binary?
	\item How prevalent is the use of virtual inheritance in MSVC binaries?
	\item How does the presence of virtual inheritance reduce the effectiveness of state of the art binary level defenses like Marx? 
	\item How accurately can the overall class inheritance (single, multiple and virtual) tree be recovered?
\end{itemize}

All ELF binaries were compiled with GCC 7.3.0 (clang+llvm 7.0) on Ubuntu 18.04 LTS, whereas Windows PE binaries were compiled using Visual Studio on Windows 10 OS. All experiments were running on Intel
core i7 3.60GHz with 32GB RAM. We compared the results from our analysis with the ground truth. Ground truth (GT) for all the binaries except mysqld, mysqlbinlog and mysqlpump were obtained from GCC's compilation option -fdump-class-hierarchy, which dumps a representation of the hierarchy of each class, including their VTable layout~\cite{gcc-fdump} and VTTs. Mysqld, mysqlbinlog and mysqlpump are together in a single package, therefore, to know their distinct inheritance trees, we analyzed RTTI structures in their binaries.

\ignore{
	\paragraph{Binaries Evaluated} We only evaluated some of the programs we found to have virtual inheritance (Table~\ref{vh_distr}) because the other binaries have their virtual bases as imported classes as such the vptrs found in their base class constructors point to the extern section of the binary. We cannot verify vptrs pointing to the extern section since the actual VTable is not in the binary, that makes it impossible for us to identity any relationship in that case. There are also certain base classes whose regular VTables are not in the binary, but have one or more construction VTables. We use one of the construction VTables to represent such classes. Note that this is needed only to construct the hierarchy. Since we know which VTables are construction VTables, they can be easily removed from the class hierarchy in order to enforce a security policy.
}
\subsection{VTT Recovery}
We report the number of distinct VTTs recovered from binaries in Table~\ref{no_vtt}. VTTs are reliable indication of virtual inheritance in a given binary. 
VTTs could also be used as a basis for comparing two binaries for similarity. Recovering a sufficient number of VTTs from two binaries and analyzing their entries can indicate if they are similar or not. Lastly, VTTs can be reliably used to verify VTables, and to differentiate them from construction VTables. As shown in the table, the number of VTTs recovered range from 1 to 166.

\begin{table}[ht]
	\centering
	\small
	\caption{Number of VTTs recovered.}
	\label{no_vtt}
	\rowcolors{1}{white}{gray!30}
	\begin{tabular}{ll}
		\toprule[0.4ex]
		\textbf{Program} & \textbf{No of VTTs} \\ \midrule
		\specialcell{libabw, boost\_date\_time, \\libcdr, libgdcmMEXD, libGLU,\\ libopencv\_phase\_unwrapping,\\ librevenge-generators,\\ librevenge-stream} & 1 \\ 
		\specialcell{libgdcmDSED, libopencv\_features2d,\\ libopencv\_structured\_light,\\ libphonenumber, VBoxRT} & 2 \\ 
		\specialcell{libepub-gen, libetonyek,\\ librados, libglibmm-2.4} & 3 \\ 
		libsocket++ & 4 \\ 
		boost\_iostream, boost\_locale, libgdal & 5 \\ 
		libcmis-c, libstorelo & 6 \\ 
		boost\_thread & 8 \\ 
		libopencv\_saliency & 9 \\ 
		libstdc++ & 27 \\
		\midrule
		\specialcell{cmake, ctest, cpack,\\ btag, k4dirstat, kgeography,\\ scantailor} & 1 \\ 
		bedtools, between, & 2 \\ 
		grfcodec, primrose & 3 \\ 
		gpick, xboxdrv, mysqlbinlog & 6 \\ 
		fityk & 7 \\ 
		mysqlpump & 10 \\
		x86\_64-linux-gnu-ld.gold & 11 \\ 
		x86\_64-linux-gnu-dwp & 12 \\ 
		darkice & 22 \\ 
		ragel & 45 \\
		Mysqld & 166 \\
		\bottomrule[0.4ex]
	\end{tabular}
\end{table}

\subsection{Virtual Inheritance Recovery}
Table~\ref{vbase_eval_lib} shows our analysis result for virtual inheritance compared with the ground truth. First, we identified classes which have at least one virtual base. After that, the number of direct bases and number of intermediate bases which those classes have was counted. It is possible to have a class with only virtual bases and no intermediate, that is the reason we have a column "0" under "Intermediate bases". We used "\textgreater1" to denote all other numbers of virtual and intermediate bases because most classes have 2 or less numbers of virtual and intermediate bases. The compiler may choose to eliminate an entire class from the compiled binary, for instance, because such class is not initialized in the program. We removed such classes from the GT used for comparison.

For libraries, we correctly recovered 95\% of virtual bases and 95.5\% of intermediate bases. We underestimated 5\% of virtual bases and 4.5\% of intermediate bases. \ignore{We missed some classes because they are not present in the binary, the binary was manually checked to confirm this.} For executables, we correctly recovered 98.8\% of virtual bases and underestimated 1.2\% of them, also, we correctly recovered 76\% of intermediate bases and underestimated 24\%. In DealII, ``LaplaceSolver::PrimalSolver\textless3\textgreater" (LP) has ``LaplaceSolver::Solver\textless3\textgreater" (LS) as its intermediate base, but the main constructor (or destructor) of LP where the VTT entry of LS should be initialized is not present in the binary. As a result we missed the relationship between LS and LP. Five of the six classes with underestimated intermediate bases have LP as intermediate base, while LP is the sixth class. This is the case for most of the underestimations. No overestimation was recorded.

\begin{table*}[ht]
	\centering
	\caption{Table showing the \# of classes in virtual inheritance tree. The ``\#Classes with virt inh" are classes with atleast 1 direct or indirect virtual base, the ``0" subcolumn intermediate bases implies only direct bases\ignore{(``0" is not under ``Virtual Bases" because a class must have at least a virtual base to be in a virtual inheritance tree)} , ``\#matching" represents the \# of bases that were correctly recovered, ``\#over-est" and ``\#under-est" imply that we recovered more and less bases than the GT respectively and ``\#not found" are the virtual inheritance instances we missed. \ignore{``Virtual Bases" show the number virtual bases each recovered class has. Columns ``Intermediate Bases" show the number of intermediate bases each of the recovered class has.} Under ``Intermediate bases", some entries have ``N/A", this is because those programs have no class with an intermediate base.}
	\label{vbase_eval_lib}
	\scalebox{0.8}{%
		\begin{tabular}{lccccccccccccc}
			\toprule[0.5ex]
			\multirow{2}{*}{\textbf{Program}} & \multirow{2}{*}{\textbf{\specialcell{\#Classes with\\ virt inh}}} & \multicolumn{5}{c}{\textbf{Virtual Bases}} & \multicolumn{6}{c}{\textbf{Intermediate bases}} & \multirow{2}{*}{\textbf{\#not found}}  \\ \cmidrule(r){3-7} \cmidrule(r){8-12}
			&  & \textbf{1} & \textbf{\textgreater{}1} & \textbf{\#matching} & \textbf{\#overest} & \textbf{\#underest} & \textbf{0} & \textbf{1} & \textbf{\textgreater{}1} & \textbf{\#matching} & \textbf{\#overest} & \textbf{\#underest} &   \\ \midrule
			libstdc++6 & 29 & 29 & 0 & 29 & 0 & 0 & 8 & 19 & 2 & 27 & 0 & 2 & 1  \\
			libcmis-c & 6 & 6 & 0 & 6 & 0 & 0 & 6 & 0 & 0 & N/A & N/A & N/A & 5  \\
			libcmis & 26 & 26 & 0 & 26 & 0 & 0 & 26 & 0 & 10 & 26 & 0 & 0 & 6  \\
			libcdr & 2 & 2 & 0 & 2 & 0 & 0 & 2 & 0 & 0 & N/A & N/A & N/A & 0  \\
			libepub-gen & 3 & 3 & 0 & 1 & 0 & 0 & 3 & 0 & 0 & N/A & N/A & N/A & 1  \\
			libetonyek & 3 & 3 & 0 & 3 & 0 & 0 & 3 & 0 & 0 & N/A & N/A & N/A & 0  \\
			libgdal & 15 & 15 & 0 & 13 & 0 & 2 & 1 & 14 & 0 & 14 & 0 & 1 & 3  \\
			librados & 3 & 3 & 0 & 3 & 0 & 0 & 3 & 0 & 0 & N/A & N/A & N/A & 0 \\ \midrule
			mysqld & 201 & 191 & 10 & 201 & 0 & 0 & 65 & 86 & 50 & 200 & 0 & 1 & 0  \\
			mysqlbinlog & 16 & 16 & 0 & 16 & 0 & 0 & 10 & 0 & 6 & 16 & 0 & 0 & 0  \\
			mysqlpump & 26 & 26 & 0 & 26 & 0 & 0 & 15 & 1 & 10 & 26 & 0 & 0 & 0  \\
			DealII & 8 & 8 & 0 & 8 & 0 & 0 & 4 & 3 & 1 & 8 & 0 & 6 & 1  \\
			Ragel & 47 & 47 & 0 & 46 & 0 & 1 & 13 & 6 & 28 & 25 & 0 & 22 & 0  \\
			Darkice & 14 & 4 & 10 & 12 & 0 & 2 & 3 & 10 & 1 & 5 & 0 & 9 & 8  \\
			
			Between & 2 & 2 & 0 & 2 & 0 & 0 & 1 & 1 & 0 & 2 & 0 & 0 & 0  \\
			Btag & 1 & 1 & 0 & 1 & 0 & 0 & 1 & 0 & 0 & N/A & N/A & N/A & 0  \\
			
			gpick & 5 & 5 & 0 & 5 & 0 & 0 & 5 & 0 & 0 & N/A & N/A & N/A & 3  \\
			grfcodec & 3 & 3 & 0 & 3 & 0 & 0 & 3 & 0 & 0 & 3 & 0 & 0 & 4  \\
			
			primrose & 3 & 3 & 0 & 3 & 0 & 0 & 1 & 2 & 0 & 3 & 0 & 0 & 0  \\
			Scantailor & 1 & 1 & 0 & 1 & 0 & 0 & 1 & 0 & 0 & N/A & N/A & N/A & 0  \\
			xboxdrv & 5 & 5 & 0 & 5 & 0 & 0 & 5 & 0 & 0 & N/A & N/A & N/A & 1  \\ 
			\bottomrule[0.5ex]
		\end{tabular}
	}
\end{table*}

\subsection{Recovery for Higher Levels of Optimization}
In order to ascertain that \codename\ performs effectively with higher levels of optimization, we compiled the library test set with the default compiler flags specified by the library authors in the configuration files. We did not alter the default configurations which we verified to be O2 optimization for all the libraries. This approach of evaluating with default compiler options is consistent with existing works~\cite{marx:andre, ooanalyzer:2018}. This evaluation is done by compiling with both GCC and Clang (Table~\ref{gcc-O2}). As expected, with higher levels of optimization some constructs (e.g. VTables) needed for recovery are missing in this binary. This is not a limitation of our tool because the needed information is infact not in the binary. For GCC, we correctly recovered 83.3\% and 82.9\% of virtual and intermediate bases. We underestimated 16.7\% and 17.1\% while we overestimated 3.6\% and 1.4\% of virtual and intermediate bases respectively. For Clang binaries, we correctly recovered 77.5\% and 73.8 of  virtual and intermediate bases. We underestimated 22.5\% and 26.2\% while we overestimated 2.5\% and 0\% of virtual and intermediate bases respectively. We were unable to compile libstdc++6 with clang, while the other 5 binaries(e.g libcmis-c) which have smaller numbers virtual inheritance occurrences do not contain any VTT when compiled with clang, default optimization.
\begin{table*}[ht]
	\centering
	\caption{Library test set compiled with higher levels of optimization using both GCC and Clang}
	\label{gcc-O2}
	\scalebox{0.75}{%
		\begin{tabular}{llccccccccccccc}
			\toprule[0.5ex]
			\multirow{2}{*}{\textbf{Compiler}} &\multirow{2}{*}{\textbf{Program}} & \multirow{2}{*}{\textbf{\specialcell{\#Classes with\\ virt inh}}} & \multicolumn{5}{c}{\textbf{Virtual Bases}} & \multicolumn{6}{c}{\textbf{Intermediate bases}} & \multirow{2}{*}{\textbf{\#not found}} \\ \cmidrule(r){4-8} \cmidrule(r){9-13}
			& &  & \textbf{1} & \textbf{\textgreater{}1} & \textbf{\#matching} & \textbf{\#overest} & \textbf{\#underest} & \textbf{0} & \textbf{1} & \textbf{\textgreater{}1} & \textbf{\#matching} & \textbf{\#overest} & \textbf{\#underest} &   \\ \midrule
			\multirow{7}{*}{\textbf{GCC}} &libstdc++6 & 26 & 26 & 0 & 26 & 0 & 0 & 4 & 17 & 5 & 25 & 0 & 0 & 4  \\
			&libcmis-c & 6 & 6 & 0 & 6 & 0 & 0 & 6 & 0 & 0 &N/A  & N/A & N/A & 5  \\
			&libcmis & 20 & 17 & 3 & 17 & 3 & 0 & 11 & 1 & 8 & 19 & 1 & 0 & 12  \\
			&libcdr & 1 & 1 & 0 & 1 & 0 & 0 & 1 & 0 & 0 &N/A  & N/A & N/A& 1  \\
			&libepub-gen & 3 & 3 & 0 & 1 & 0 & 0 & 3 & 0 & 0 &N/A  & N/A & N/A&1    \\
			&libetonyek & 3 & 3 & 0 & 3 & 0 & 0 & 3 & 0 & 0 &N/A  & N/A & N/A& 0  \\
			&libgdal & 15 & 15 & 0 & 13 & 0 & 2 & 1 & 14 & 0 & 14 & 0 & 1 & 3  \\ 
			\midrule
			\multirow{2}{*}{\textbf{Clang}}& libcmis & 17 & 17 & 0 & 17 & 0 & 0 & 7 & 0 & 10 &17  & 0 & 0& 15  \\
			&libgdal & 16 & 14 & 2 & 14 & 2 & 0 & 1 & 15 & 0 & 14 & 0 & 2 & 2  \\ 
			\bottomrule[0.4ex]
	\end{tabular}}
\end{table*}

\subsection{MSVC Binaries}
\begin{figure}
	\centering
	\includegraphics[height=4cm, width=8cm]{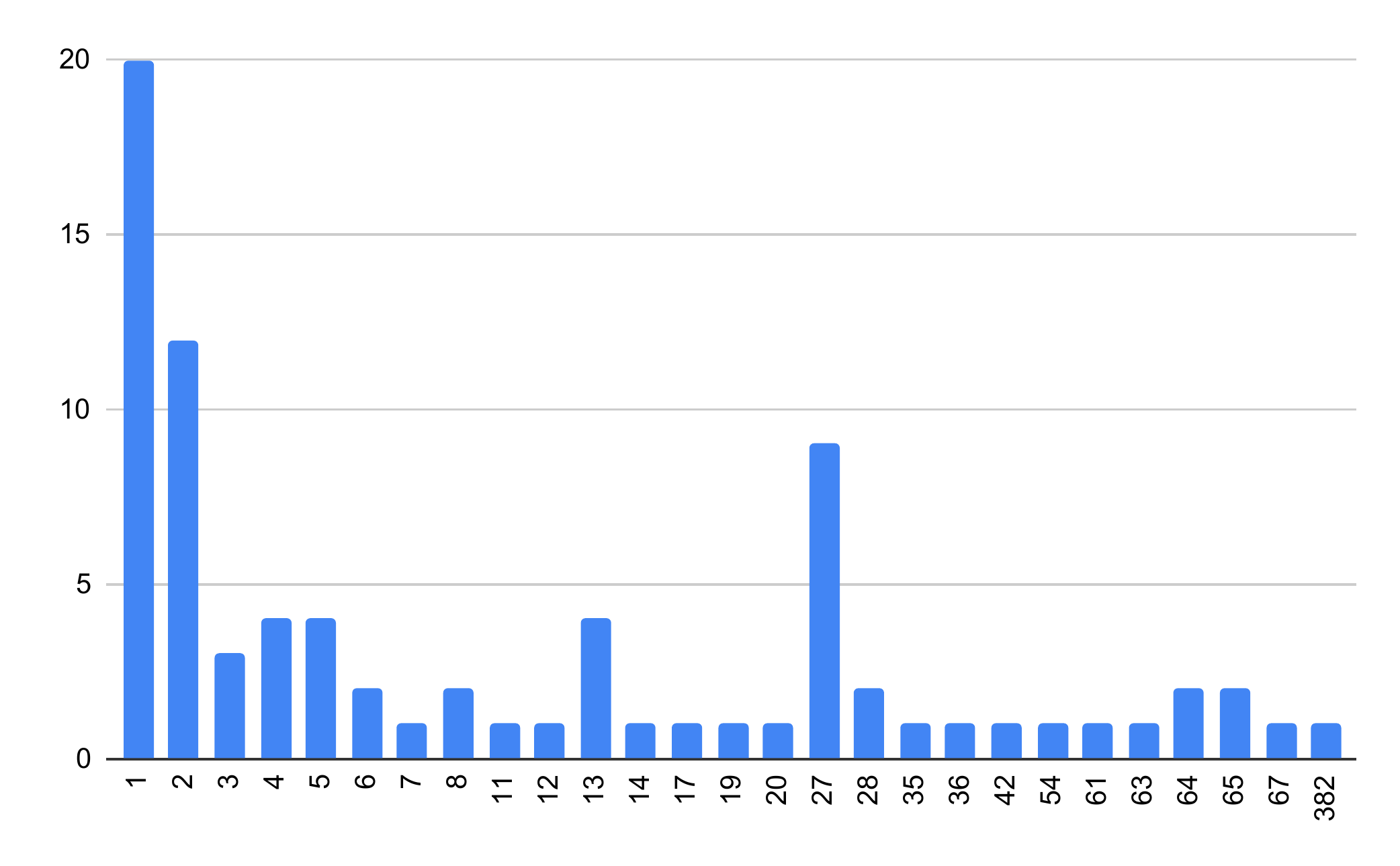}
	\caption{MSVC: No of classes with virt inheritance}\label{fig:msvc}
\end{figure}

We analyzed 648 DLLs which we recovered from various directories (including Program Files) of Windows machine. The aim of this evaluation is to find out how prevalent the use of virtual inheritance is among Windows applications. All these DLLs contain polymorphic classes. Of the 648 DLLs, we found 81 (12.5\%) to contain instances of virtual inheritance. Table~\ref{msvc_5} shows the DLLs with the top 15 number of classes with virtual inheritance. We manually checked the binaries to verify that the reported numbers are true positives. Figure~\ref{fig:msvc} shows the distribution of the number of classes with virtual inheritance among the 81 DLLs. The x-axis shows the number of classes while the y-axis shows the number of DLLs with those numbers.

\begin{table}[]
	\centering
	\caption{MSVC Binaries (Top 5 with virtual inheritance)}
	\label{msvc_5}
	\begin{tabular}{llll}
		\toprule[0.4ex]
		\multirow{2}{*}{DLLs} & \multicolumn{3}{c}{Virtual Bases} \\ \cmidrule(r){2-4}
		& \#classes with virt inh & 1 & \textgreater{}1 \\ \hline
		migcore & 382 & 292 & 90 \\
		igd11dxva32 & 67 & 2 & 65 \\
		igd9dxva32 & 65 & 5 & 60 \\
		igd12dxva32 & 65 & 5 & 60 \\
		igd9dxva64 & 64 & 37 & 27 \\
			igd11dxva64 & 64 & 37 & 27 \\
			igd12dxva64 & 63 & 36 & 27 \\
			MSVidCtl & 61 & 43 & 18 \\
			msvcp60 & 54 & 2 & 52 \\
			igc32 & 42 & 1 & 41 \\
			migstore & 36 & 28 & 8 \\
			intelocl64 & 35 & 32 & 3 \\
			BingMaps & 28 & 15 & 13 \\
			igdrcl64 & 28 & 2 & 26 \\
			csiagent & 27 & 26 & 1 \\ 
		\bottomrule[0.4ex]
	\end{tabular}
\end{table}

	\subsection{Overall Class Inheritance Recovery}
	Table ~\ref{overall_hierarchy} shows our result for the overall class inheritance recovery compared with the ground truth. The table shows the number of classes, most base classes (classes with no base), classes with single base, classes with multiple bases and classes in virtual inheritance tree for GT and \codename. We recorded an average precision of 99.8\% and 99.3\% overall hierarchy for libraries and executables respectively. Compared to the total class hierarchy from the ground truth, we recovered an average of 66.5\% and 82\% for libraries and executables respectively. Most of the missing classes come from Boost libraries.\ignore{\ayomide{find out why classes from boost are missing}}
	
	\begin{table*}[]
		\centering
		\caption{Table showing the overall class hierarchy from the analysis compared to the ground truth for both libraries and executables}
		\label{overall_hierarchy}
		\begin{tabular}{lcccccccccc}
			\toprule[0.5ex]
			\multicolumn{1}{c}{\multirow{2}{*}{\textbf{Program}}} & \multicolumn{2}{c}{\textbf{\#Classes}} & \multicolumn{2}{c}{\textbf{\#most base classes}} & \multicolumn{2}{c}{\textbf{\# with single base}} & \multicolumn{2}{c}{\textbf{\#with multiple bases}} & \multicolumn{2}{c}{\textbf{\#with virt inh}} \\
			\cmidrule(r){2-3} \cmidrule(r){4-5} \cmidrule(r){6-7} \cmidrule(r){8-9} \cmidrule(r){10-11}
			\multicolumn{1}{c}{} & \textbf{GT} & \textbf{Analysis} & \textbf{GT} & \textbf{Analysis} & \textbf{GT} & \textbf{Analysis} & \textbf{GT} & \textbf{Analysis} & \textbf{GT} & \textbf{Analysis} \\
			\midrule
			libstdc++ & 211 & 202 & 11 & 12 & 172 & 164 & 1 & 0 & 27 & 26 \\
			libcmis-c & 78 & 30 & 39 & 11 & 21 & 10 & 7 & 3 & 11 & 6 \\
			libcmis & 230 & 194 & 54 & 28 & 134 & 133 & 13 & 7 & 32 & 26 \\
			libcdr & 181 & 57 & 48 & 20 & 69 & 33 & 6 & 2 & 58 & 2 \\
			libepub-gen & 90 & 77 & 27 & 15 & 59 & 59 & 2 & 2 & 2 & 3 \\
			libetonyek & 1008 & 814 & 50 & 42 & 884 & 768 & 17 & 1 & 57 & 3 \\
			libgdal & 1103 & 1024 & 131 & 146 & 948 & 855 & 6 & 8 & 18 & 15 \\
			librados & 213 & 120 & 86 & 56 & 240 & 58 & 13 & 3 & 12 & 3 \\
			\midrule
			mysqld & 3640 & 3666 & 252 & 408 & 3144 & 3010 & 46 & 47 & 198 & 201 \\
			mysqlbinlog & 66 & 71 & 5 & 15 & 30 & 24 & 15 & 16 & 16 & 16 \\
			mysqlpump & 117 & 123 & 10 & 19 & 77 & 74 & 4 & 4 & 26 & 26 \\
			DealII & 874 & 711 & 25 & 22 & 836 & 678 & 4 & 3 & 9 & 8 \\
			ragel & 74 & 73 & 3 & 2 & 24 & 24 & 0 & 0 & 47 & 47 \\
			Darkice & 32 & 29 & 10 & 13 & 0 & 0 & 0 & 0 & 22 & 16 \\
			Between & 47 & 45 & 15 & 16 & 27 & 24 & 3 & 3 & 2 & 2 \\
			
			gpick & 82 & 63 & 25 & 16 & 43 & 38 & 6 & 4 & 8 & 5 \\
			grfcodec & 45 & 30 & 18 & 12 & 11 & 16 & 7 & 4 & 7 & 3 \\
			primrose & 28 & 27 & 7 & 7 & 18 & 17 & 0 & 0 & 3 & 3 \\
			xboxdrv & 156 & 150 & 22 & 23 & 118 & 119 & 6 & 4 & 6 & 5\\ \bottomrule[0.4ex]
			
		\end{tabular}
	\end{table*}

\subsection{Comparison with Marx}
We compare \codename\ with Marx, by considering the number of classes in virtual inheritance tree which Marx and \codename\ recovered. We attempted to do similar comparisons with VCI and SmartDec. However, VCI is not open sourced and the authors did not release the source code to us. We compiled a version of SmartDec that we found on GitHub, but the tool does not do what the paper describes. It tries to recover source code from a binary rather than recover class hierarchy. Marx groups classes into sets while we assign direction of inheritance to every class. To achieve a fair comparison, we evaluated the number of distinct virtual inheritance trees which we recovered. Column "\#cvtables" shows the number of construction VTables which Marx groups with regular VTables (these constitute false positives). For all the binaries, except libstdc++ and libetonyek, Marx groups classes involved in virtual inheritance into 1 or 2 sets. These sets contain the virtual bases, intermediate bases and other classes with either single or multiple inheritance. For libstdc++, Marx groups each class with virtual inheritance into separate sets with no other class in them. None of the sets contain either virtual or intermediate bases, they were missed. For libetonyek, Marx groups classes involved in virtual inheritance into 3 separate sets. Those sets also contain other classes not involved in virtual inheritance with neither the virtual or intermediate bases being present. Lastly, Marx does not reason about virtual inheritance, as a result column "\#Edges in VIT" is zero for all binaries. \ignore{In Figure~\ref{cvtable-distr} we show the distribution of vbase-offset and offset-to-top values found in the construction VTables of the programs represented. As discussed in Section~\ref{sec-impl}, these offsets can be easily exploited by an attacker to achieve a data corruption attack.}

\begin{table*}[ht]
	\centering
	\caption{Table comparing the representation of Marx for class involved in virtual inheritance with \codename\ for both libraries and executables. ``\#set" shows the number of sets recovered by Marx containing at least one class with a virtual base class. ``\#classes in set" shows the total number of classes in the sets. ``\#cvtables" is the total number of constructor VTables wrongly identified as regular VTables. ``\#Edges in VIT" shows number of edges in Virtual Inheritance Tree}
	\label{comp_with_marx2}
	\scalebox{0.9}{%
		\begin{tabular}{lllllllll}
			\toprule[0.6ex]
			\footnotesize
			\multirow{2}{*}{\textbf{Program}} & \multicolumn{4}{c}{\textbf{Marx}} & \multicolumn{4}{c}{\textbf{Our analysis}} \\
			\cmidrule(r){2-5} \cmidrule(r){6-9}
			& \textbf{\#set} & \textbf{\specialcell{\# classes \\in set}} & \textbf{\specialcell{\#cvtables\\(falses)}} & \textbf{\specialcell{\#Edges in VIT}} & \textbf{\#tree} & \textbf{\#in tree} & \textbf{\specialcell{\#cvtables\\(falses)}} & \textbf{\specialcell{\#Edges in VIT}} \\
			\midrule
			libstdc++ & 26 & 32 & 2 & 0 & 2 & 30 & 0 & 31 \\
			libcmis-c & 1 & 7 & 0 & 0 & 1 & 7 & 0 & 6 \\
			libcmis & 2 & 48 & 20 & 0 & 2 & 28 & 0 & 36 \\
			libcdr & 1 & 3 & 0 & 0 & 1 & 3 & 0 & 2 \\
			libepub-gen & 1 & 4 & 0 & 0 & 1 & 4 & 0 & 3 \\
			libetonyek & 3 & 4 & 0 & 0 & 1 & 4 & 0 & 3 \\
			libgdal & 1 & 31 & 15 & 0 & 1 & 16 & 0 & 15 \\
			librados & 1 & 4 & 0 & 0 & 1 & 4 & 0 & 3 \\
			\midrule
			mysqld & 8 & 271 & 69 & 0 & 6 & 202 & 0 & 259 \\
			mysqlbinlog & 1 & 30 & 10 & 0 & 4 & 20 & 0 & 22 \\
			mysqlpump & 3 & 41 & 15 & 0 & 2 & 26 & 0 & 59 \\
			DealII & 2 & 18 & 10 & 0 & 2 & 10 & 0 & 9 \\
			ragel & 1 & 100 & 50 & 0 & 3 & 50 & 0 & 70 \\
			Darkice & 1 & 23 & 4 & 0 & 1 & 19 & 0 & 32 \\
			Between & 1 & 3 & 0 & 0 & 1 & 3 & 0 & 2 \\
			gpick & 1 & 6 & 0 & 0 & 1 & 6 & 0 & 5 \\
			grfcodec & 1 & 4 & 0 & 0 & 1 & 4 & 0 & 3 \\
			primrose & 1 & 4 & 0 & 0 & 1 & 4 & 0 & 3 \\
			xboxdrv & 1 & 6 & 0 & 0 & 1 & 6 & 0 & 5 \\
			\bottomrule[0.5ex]
		\end{tabular}
	}
\end{table*}

\section{Related Work}\label{sec:related}
VCI, MARX and SmartDec focus on recovering single and multiple inheritance from a C++ binary. VCI analyses constructors to recover single and multiple inheritance tree. The constructor of a derived class calls the constructors of its base classes, therefore VCI uses this order information to identify base classes of a derived class. It does not consider virtual inheritance. 

Marx is slightly similar to VCI, however, it uses a more intuitive approach. During calls to base class constructors or destructors, the vptr within the derived class object gets overwritten. Only vptrs of related classes can be overwritten. Marx analyses vptr overwrites to group related classes. Its weakness is in its inability to differentiate between a constructor and destructor which makes it impossible to assign direction of inheritance. Marx also does not reason about virtual inheritance, it simply groups vptrs for construction VTables and complete object VTables together. 

SmartDec considers both constructor and destructors to recover inheritance. Like VCI, it identifies the base classes by considering the calls to constructors and destructors in the derived class' constructor and destructor respectively. It only considers single and multiple inheritance, not virtual inheritance.

OOAnalyzer takes a different approach in recovering high level semantics in C++ programs. It does not rely on VTables which makes it possible to consider both polymorphic and non-polymorphic classes. However, this makes it hard to recover inheritance since vptr initialization is a strong indication of class relationships.

Katz et al.\cite{katz:2016} proposed an approach to statically determine the possible targets of virtual function calls. This is achieved by first identifying object tracelets, a statically constructed sequence of operations performed on an object. These object tracelets are then used to train a statistical language model (SLM) for every type. The resulting ensemble of SLMs is used to generate a ranking of their most likely types, from which the likely targets of dynamic dispatches are deduced. Basically, the ensemble of SLMs is used to measure the likelihood that sets of tracelets share the same source, those set of tracelets are grouped together, which then form the basis for predicting possible targets of virtual function calls. The grouping of object types is similar to what Marx does.

\codenametwo~\cite{declassifier} implements several techniques to recover class hierarchy information from optimized binaries. These techniques include constructor/destructor analysis, overwrite analysis and object layout analysis. It achieves a high precision and accuracy on optimized binaries, however, it also does not consider virtual inheritance.

\ignore{SECRET~\cite{secret} enforces the ``What-You-Target-Is-Not-What-You-Execute" (WYTINWYX) policy to thwart attacks that depend on code pointer harvesting. To build this policy, it recovers several high level information from the binary such as VTables, jump tables, code pointer constants and so on.
	
	VTPin~\cite{vtpin} ensures that the vptr written into an object through the compiler code does not get overwritten at runtime. This is to defend against VTable hijacking attacks caused by use-after-free vulnerabilities. VTPin recovers RTTI information from the binary to validate a vptr and then pins it so that it does not get overwritten.
	
	VTint~\cite{zhang:2015:vtint} protects against VTable hijacking attacks. It first recovers VTables and virtual function dispatches in the binary, and then instruments the binary with security checks to ensure the integrity of VTables used at every virtual function dispatch.}

vfGuard~\cite{prakash:2015:vfguard} is a binary level defense that protects virtual function callsites. It statically analyzes the binary to recover VTables as well as function offsets specified at callsites. It uses this information to enforce a CFI policy that restricts virtual dispatch targets to only functions at the same offset within VTables as the static offset specified at callsite.

VTable Pointer Separation (VPS)~\cite{vps:2019} is a binary level defense that implements CFIXX’s Object Type Integrity on the binary level. It performs static and dynamic analysis and runtime instrumentation to identify the exact target vptr allowable at a given virtual callsite at runtime.

\section{Conclusion}\label{sec:conclude}
Previous binary level class hierarchy recovery solutions have made no attempt to recover the virtual inheritance tree of a program. In this work, we show the security implications of the failure to include virtual hierarchy in the overall inheritance tree. We also present simple, but efficient algorithms to recover virtual inheritance in a C++ binary.

\section{ACKNOWLEDGEMENT}\label{sec:ack}
This research was supported in part by Office of Naval Research Grant \#N00014-17-1-2929, National
Science Foundation Award \#1566532, and DARPA award
\#81192. Any opinions, findings and conclusions in this paper
are those of the authors and do not necessarily reflect the
views of the funding agencies.



\end{document}